\documentclass[superscriptaddress, amsmath,amssymb,  aps,prb, twocolumn]{revtex4-2}

\usepackage[dvipsnames]{xcolor}
\usepackage[most]{tcolorbox}

\usepackage{amsmath,amssymb}
\usepackage{amsfonts} 	
\usepackage{graphicx}   
\usepackage{verbatim}   
\usepackage{textgreek}
\usepackage{lipsum}
\usepackage{color}      
\usepackage{subfigure}  
\usepackage[normalem]{ulem} 
\usepackage[super]{nth} 
\usepackage{hyperref}   
\usepackage{pbox}
\begin{document}

\title{Reconfigurable Intelligent Surface: MIMO or radiating sheet?} 

\author{Sotiris Droulias}
\email{sdroulias@unipi.gr}
\affiliation{Department of Digital Systems, University of Piraeus, Piraeus 18534, Greece.}

\author{Angeliki Alexiou}
\affiliation{Department of Digital Systems, University of Piraeus, Piraeus 18534, Greece.}

\begin{abstract}
A Reconfigurable Intelligent Surface (RIS) redirects and possibly modifies the properties of incident waves, with the aim to restore non-line-of-sight communication links. Composed of elementary scatterers, the RIS has been so far treated as a collection of point scatterers with properties similar to antennas in an equivalent massive MIMO communication link. Despite the discrete nature of the RIS, current design approaches often treat the RIS as a continuous radiating surface, which is subsequently discretized. Here, we investigate the connection between the two approaches in an attempt to bridge the two different perspectives. We analytically find the factor that renders the two approaches equivalent and we demonstrate our findings with examples of RIS elements modeled as antennas with commonly used radiation patterns and properties consistent with antenna theory. We analyze the equivalence between the two theoretical approaches with respect to design aspects of the RIS elements, such as gain, directivity and coupling between elements, with the aim to provide insight into the observed discrepancies, the understanding of which is crucial for assessing the RIS efficiency. 
\end{abstract}


\maketitle
\section{Introduction}
\begin{figure}[b!]
\centering
		\includegraphics[width=1\linewidth]{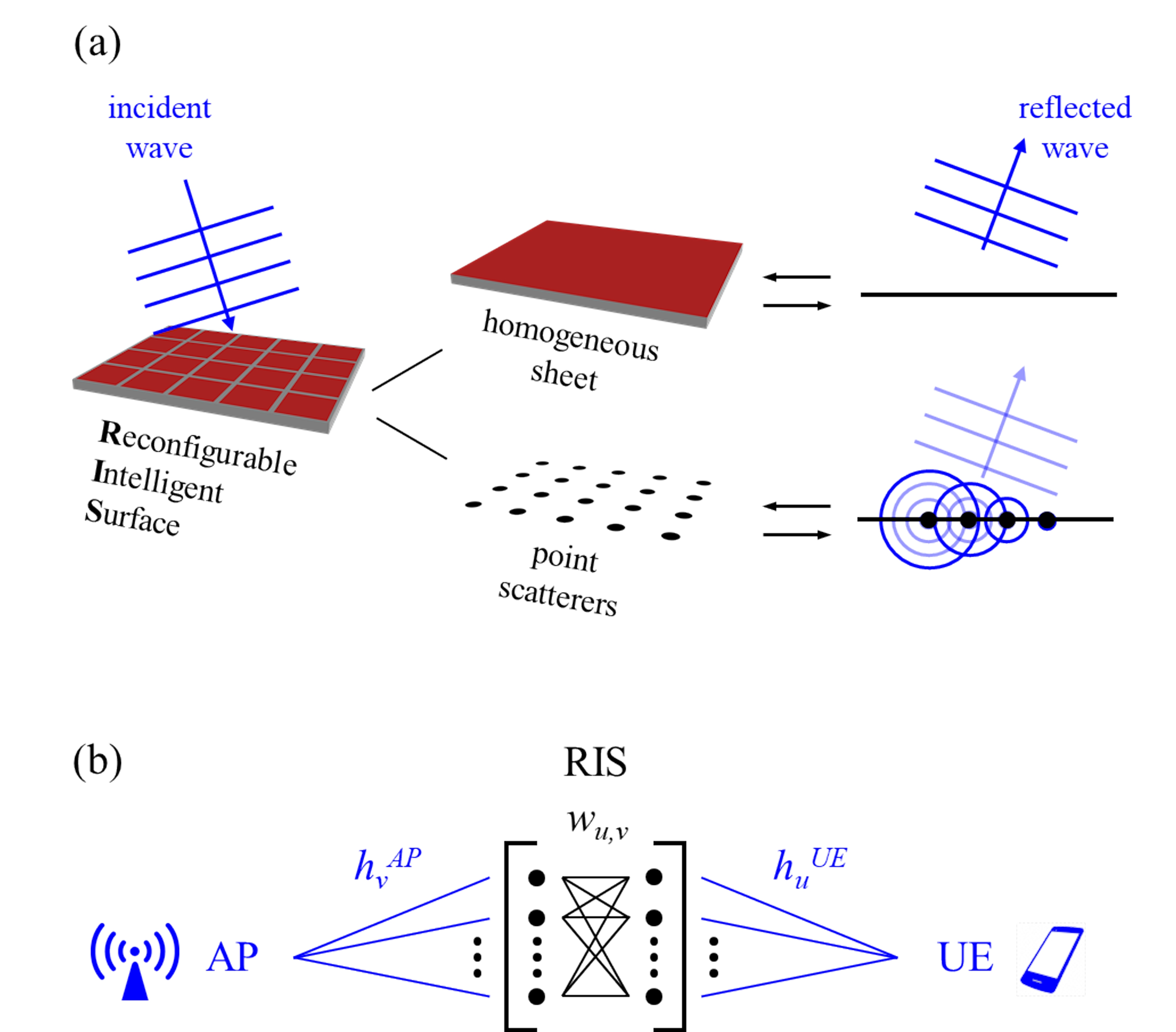}
	\caption{Schematic representation of a RIS steering an incident wave to a prescribed direction. (a) The RIS can be analyzed as a homogeneous radiating sheet or, equivalently, as a distribution of point scatterers. (b) The discrete RIS approach as an equivalent MIMO system. The AP excites the RIS elements (point scatterers), which may interact with each other via the matrix elements $w_{u,v}$ and subsequently re-radiate. The free-space path loss from the AP to the $v^{th}$ RIS element and from the $u^{th}$ RIS element to the UE is described by the matrix elements $h_v^{AP}$ and $h_u^{UE}$, respectively.
}
    	\label{fig:fig01}
\end{figure}

\noindent The primary role of a reconfigurable intelligent surface (RIS) is to mediate a non-line-of-sight link by redirecting the incident beam from the transmitter to the receiver. Its operation is similar to that of a mirror, however the reflection is not limited to specular; depending on the properties of the RIS elements (periodically distributed scatterers forming the RIS surface), the RIS may redirect the incident beam towards any desired direction and, possibly, modify its characteristics, in order to optimize the beamforming efficiency and to maximize the signal power at the receiver. In recent years, there has been a considerable effort to incorporate the functionalities offered by RISs in millimeter-wave (mmWave) (30-100 GHz) and terahertz (THz) band (0.1-10 THz) communications \cite{Epstein2016, Withayachumnankul2018, Tretyakov2020, Zhang2021}. \\
\indent So far, several techniques have been theoretically proposed for the design of the desired RIS properties \cite{Capasso2011, Eleftheriades2014, Asadchy2016, Estakhri2016, Eleftheriades2016, Radi2017, Asadchy2017, Rubio2017, Eleftheriades2018, Rubio2019, Grbic2020, Rubio2021, DiRenzo2020, Alouini2021, Matthaiou2021} and relevant experiments have been performed in order to verify the predicted RIS performance \cite{Asadchy2017, Rubio2017, Eleftheriades2018, Rubio2019, Yang2016, Zhang2018, Dai2020, Bjornson2021, DeRosny2021}. To achieve the desired wave manipulation, the design involves the determination of the appropriate surface properties, such as surface impedance (or effective electric and magnetic surface conductivities). Usually, the RIS is theoretically treated as a continuous surface, i.e. as a radiating sheet that locally satisfies the boundary conditions, ensuring that an incident plane wave is reflected towards the desired direction, as shown in Fig.\,\ref{fig:fig01}(a). The solution, which is exact for surfaces of infinite extent, leads to the prescription of a continuous local wave impedance at the RIS surface. Ideally, a continuous surface of finite extent characterized by the prescribed impedance will steer the incident wave towards the desired direction, however for the sake of practical implementation the surface must be discretized, in essence rendering the continuous surface a collection of discrete scatterers. \\
%
%
%
%
\indent Alternatively, due to the inherently discrete nature of the RIS elements, it is more natural to treat the RIS as a planar distribution of scatterers. The analysis can be simplified if the RIS elements are considered as point scatterers that bear the properties of conventional antennas, such as gain and directivity \cite{Bjornson2020, Marzetta2020, Ellingson2021, DiRenzo2021, Alexiou2021, Ntontin2021}. As with antennas, the local currents that induce radiation depend on the particular design of the RIS elements. However, while actual antennas are fed directly by external currents, here the currents are excited by external waves; the local phase and amplitude of the incident waves drive the phase and amplitude, respectively, of the locally excited currents in the RIS elements. By controlling the properties of the RIS elements (e.g., by incorporating tunable resistive and reactive elements \cite{Yang2016, Zhang2018, Dai2020, Jornet2020, Venkatesh2020, Bjornson2021, DeRosny2021}), the amplitude and phase of the current oscillation at each individual scatterer can be tuned, similarly to how reflectarrays work. As a result, the entire distribution of scatterers emits radiation with a prescribed amplitude and phase, essentially re-radiating the incident wave towards the desired direction, which is macroscopically perceived as the RIS steering the incident beam. \\
\indent On the one hand, because the former of the two approaches involves a continuous radiating sheet of infinite extent that is subsequently discretized, the question is how well the local RIS elements retain the properties prescribed by the continuous infinite radiating sheet upon discretization and what is the effect of the finite RIS size. On the other hand, because the latter of the two approaches involves the collective response of individual scatterers, therefore by definition implementing a finite-sized RIS, the question is how well the properties of the individual antenna-like scatterers can reproduce the actual field radiated from the RIS. \\
\indent 
In this work, we demonstrate the equivalence between the two approaches in an attempt to bridge the two perspectives.
We analyze each approach separately and we find the connection between the two, on the basis that they must both account for elements with the same properties, i.e. polarizabilities, guaranteeing the same scattered wave. We investigate how the properties of the continuous sheet are related to the properties of point scatterers that bear characteristics consistent with the antenna theory, and we discuss commonly used models in recent theoretical works. We find that, overall, the treatment of the RIS as point scatterers may overestimate the scattered field and, therefore, a correction factor must be taken into account. By analyzing the correction factor by means of realistic radiation patterns, we demonstrate how the RIS performance is affected by design properties of the RIS elements, such as gain, directivity and inter-element coupling, and we discuss the implications on the predicted power at the receiver. 
\section{Discrete vs continuous approach}
\noindent Let us consider a RIS consisting of $N_x \times N_y$ elements, periodically arranged with periodicity $l_x$ and $l_y$ along the $x$ and $y$ directions, respectively. Under the discrete approach, the RIS is a planar distribution of $N_x \times N_y$ point scatterers that bear the properties of conventional antennas, while under the continuous approach, the RIS is a continuous rectangular surface with dimensions $L_x = N_x \times l_x$ and $L_y = N_y \times l_y$. Following both approaches, in this section, we derive the element polarizabilities, the wave scattered by the RIS, and the power received by the user.
\subsection{RIS as MIMO}
%
%
%
\begin{figure}[t!]
\centering
		\includegraphics[width=1\linewidth]{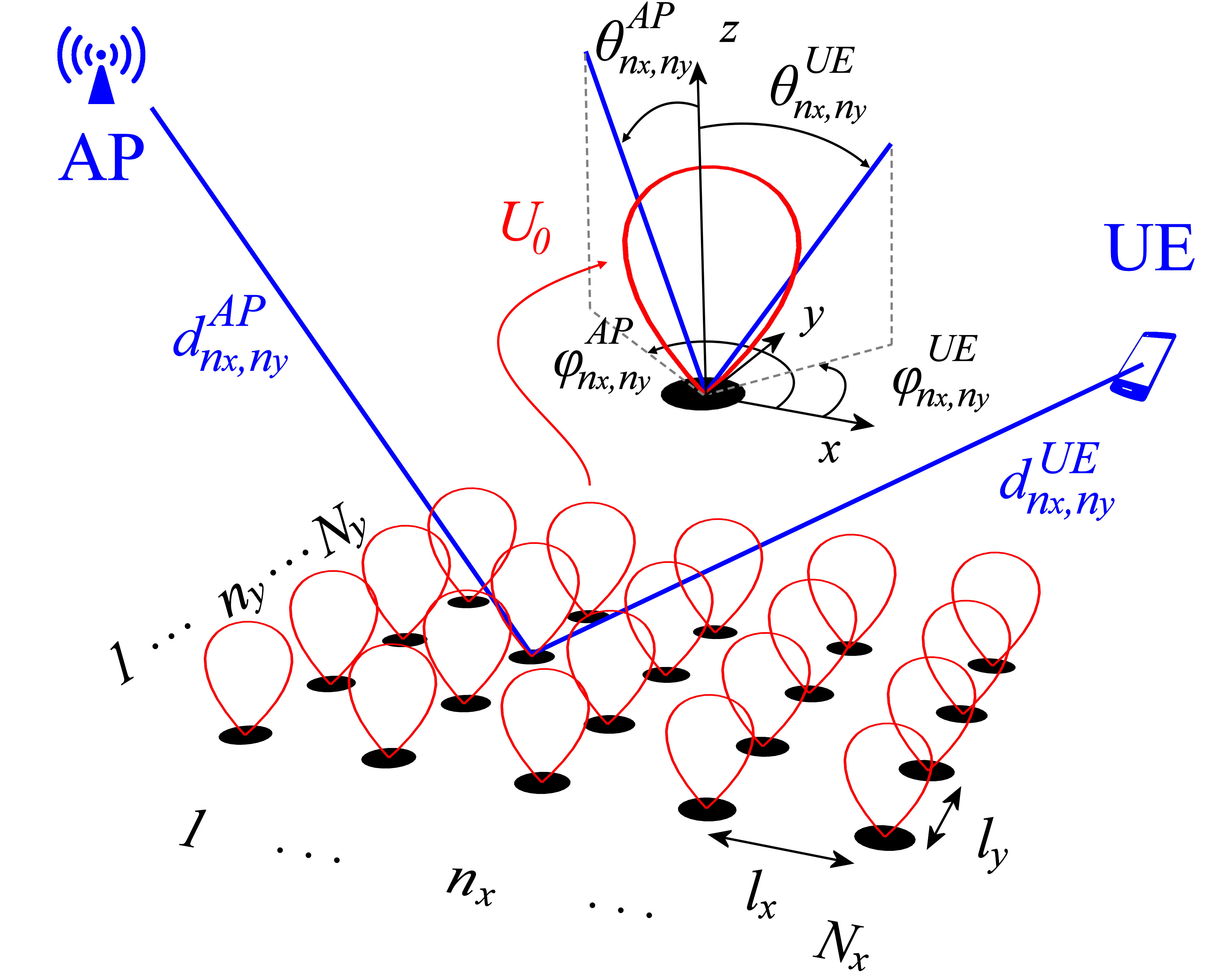}
    \caption{RIS as MIMO. The RIS is represented by a distribution of $N_x\times N_y$ point scatterers (here shown $5\times 4 = 20$ elements) periodically arranged on the $xy$-plane with periodicity $l_x$ and $l_y$ along the $x$ and $y$ axis, respectively, each characterized by the same radiation pattern, $U_0$. The angles $\theta_{n_x,n_y}^{\{AP,UE\}}$ and $\phi_{n_x,n_y}^{\{AP,UE\}}$ denote the elevation and azimuth angle, respectively, from the ($n_x,n_y$) element to the transmitter (AP) or receiver (UE) antenna, which is located at distance $d_{n_x,n_y}^{\{AP,UE\}}$.}	
    	\label{fig:fig02}
\end{figure}
%
%
\noindent Under the discrete approach, the RIS can be perceived as having $N_x\times N_y$ inputs and outputs (the same $N_x\times N_y$ elements are externally excited and subsequently re-radiate), i.e. the RIS operation can be described in terms of a matrix $\textbf{W}=(w_{u,v}) \in \mathbb{C}^{(N_x\times N_y)\times (N_x\times N_y)}$, similarly to how antennas are described in massive MIMO links:
\begin{align}
E_u^{UE} = \sum_{v=1}^{N_x\times N_y} h_u^{UE} w_{u,v} h_v^{AP} E^{AP},
    \label{Eq:EqM1}
\end{align}
where $E^{AP}$ is the field amplitude at the transmitter (or access point, AP), $h_v^{AP}E^{AP}$ is the field amplitude at the position of the $v^{th}$ element from the AP, $E_u^{UE}$ is the field amplitude at the receiver (or user equipment, UE) from the $u^{th}$ element, and $h^{UE},h^{AP}$ account for the free-space path loss between the UE and AP, respectively, and the corresponding RIS element. This is schematically shown in Fig.\,\ref{fig:fig01}(b). The off-diagonal elements of matrix $\textbf{W}$ account for possible coupling between the RIS elements, i.e. when the excited fields in one element affect the local currents of neighboring elements, particularly if the RIS elements are closely spaced. In the absence of such coupling, the matrix $\textbf{W}$ becomes diagonal, and the remaining elements can be organized so that their position $u=v\equiv (n_x,n_y)$ on the matrix can be essentially associated with their geometric location on the RIS, as illustrated in Fig.\,\ref{fig:fig02}. 
In this case, the field amplitude at the UE from the $u^{th}$ element is simply written as $E_{n_x,n_y}^{UE} = h_{n_x,n_y}^{UE} w_{n_x,n_y} h_{n_x,n_y}^{AP} E^{AP} \equiv h_{n_x,n_y}^{UE} w_{n_x,n_y} E_{n_x,n_y}^{AP}$, where $E_{n_x,n_y}^{AP}$ is the field amplitude at the position of the $(n_x,n_y)$ element from the AP.
Assuming that all elements are identical, having the same normalized radiation pattern $U_0$ and, hence, common aperture $A_{UC}$ and gain $G_{UC}$, the electric field from the $(n_x,n_y)$ RIS element to the receiver antenna, which is at distance $d_{n_x,n_y}^{UE}$, can be written in the form of dipole radiation field as (see Appendix~\ref{Sec:Appendix_A} for details):
\begin{align}
    E_{n_x,n_y}^{UE} = \underbrace{\frac{1}{4\pi \epsilon_0} \frac{e^{-j k_0 d_{n_x,n_y}^{UE}}}{d_{n_x,n_y}^{UE}}k_0^2}_{ h_{n_x,n_y}^{UE} } \underbrace{\alpha_{n_x,n_y}^{MIMO}}_{ w_{n_x,n_y} } E_{n_x,n_y}^{AP},
   \label{Eq:EqM2}
\end{align}
where
\begin{align}
    \alpha_{n_x,n_y}^{MIMO} = R_{n_x,n_y} \sqrt{A_{UC} U_{UC}^{AP} G_{UC} U_{UC}^{UE}} \frac{\sqrt{4\pi} \epsilon_0}{k_0^2}
   \label{Eq:EqM3}
\end{align}
is the polarizability of the ($n_x,n_y$) RIS element, $k_0=2\pi/\lambda_0$ ($\lambda_0$ is the free-space wavelength), and $\epsilon_0$ is the vacuum permittivity. In this notation, $U_{UC}^{\{AP,UE\}}\equiv U_0(\theta^{\{AP,UE\}}_{n_x,n_y}$, $ \phi^{\{AP,UE\}}_{n_x,n_y})$ is the common element (or unit cell, UC) radiation pattern, with $\theta_{n_x,n_y}$, $\phi_{n_x,n_y}$ denoting the elevation and azimuth angle from the ($n_x,n_y$) element to the AP and UE, as indicated in the superscript and illustrated in Fig.\,\ref{fig:fig02}. $R_{n_x,n_y}$ is a complex coefficient, with $|R_{n_x,n_y}|^2$ accounting for power loss ($|R_{n_x,n_y}|^2< 1$) or gain ($|R_{n_x,n_y}|^2> 1$) and arg$(R_{n_x,n_y})$ for the excitation phase of the ($n_x,n_y$) element. \\
\indent Using Eqs.\@(\ref{Eq:EqM2}),\@(\ref{Eq:EqM3}), the power density at the receiver can be expressed via the sum of the field contributions from all elements as $S_{r}=|\sum_{n_x}\sum_{n_y} E^{UE}_{n_x,n_y}|^2/2Z_0$ \cite{DiRenzo2021, Alexiou2021}, where $Z_0$ is the free-space wave impedance. The resulting received power $P_{r}$ is given by:
\begin{multline}
      P_{r} = A_{r} S_{r} = \frac{A_{r}}{2 Z_0} \times \\ 
\left|\sum_{n_x}\sum_{n_y} R_{n_x,n_y} \sqrt{A_{UC} U_{UC}^{AP} G_{UC} U_{UC}^{UE}} \frac{e^{-j k_0 d_{n_x,n_y}^{UE}}}{\sqrt{4\pi} d_{n_x,n_y}^{UE}} E_{n_x,n_y}^{AP}\right|^2,  
    \label{Eq:EqM4}
\end{multline}
where $A_{r}=G_{r}\lambda_0^2/4\pi$ is the aperture and $G_{r}$ is the gain of the receiver's antenna. By properly assigning the magnitude and phase of $E_{n_x,n_y}^{AP}$, the received power can be calculated under different RIS illumination conditions, such as illumination by plane waves, spherical waves \cite{Ellingson2021, DiRenzo2021, Alexiou2021, Ntontin2021} or beams of finite extent\cite{Stratidakis2021a, Stratidakis2021b, Stratidakis2022}.
\subsection{RIS as radiating sheet}
\noindent When the RIS is considered as a radiating sheet, the starting point is to represent the actual RIS by an electrically thin surface of infinite extent, which carries surface electric and magnetic currents that circulate in such a manner so that the sheet redirects incident plane waves similarly to how the RIS does. For finite-sized sheets, however, an incident plane wave is not reflected into just a single plane wave. 
To determine the scattered field, without loss of generality, we consider incident a TE-polarized wave traveling on the $xz$-plane, $\textbf{E}_i(\textbf{r}) = E_i e^{-j \textbf{k}_i \textbf{r}} \hat{\textbf{y}}$, where $\textbf{k}_i$ is the wavevector. The sheet has size $L_x \times L_y$ and steers the wave on the $xz$-plane, as shown in Fig.\,\ref{fig:fig03}. The sheet reflection coefficient associates the reflected with the incident wave at $z=0$ as $\textbf{E}_r(\textbf{r})=\Gamma_s \textbf{E}_i(\textbf{r})$ and has the form $\Gamma_s(x)=\Gamma_0 e^{j k_0(\sin{\theta_i}-\sin{\theta_r})x}$, where $\Gamma_0$ is a complex constant and $\theta_i,\theta_r$ are the angles of incidence and reflection, respectively. Following the steps outlined in Appendix~\ref{Sec:Appendix_B}, we find that the power captured by a receiver with antenna aperture $A_r$ is:
%
%
%
%
%
\begin{figure}[t!]
\centering
		\includegraphics[width=1\linewidth]{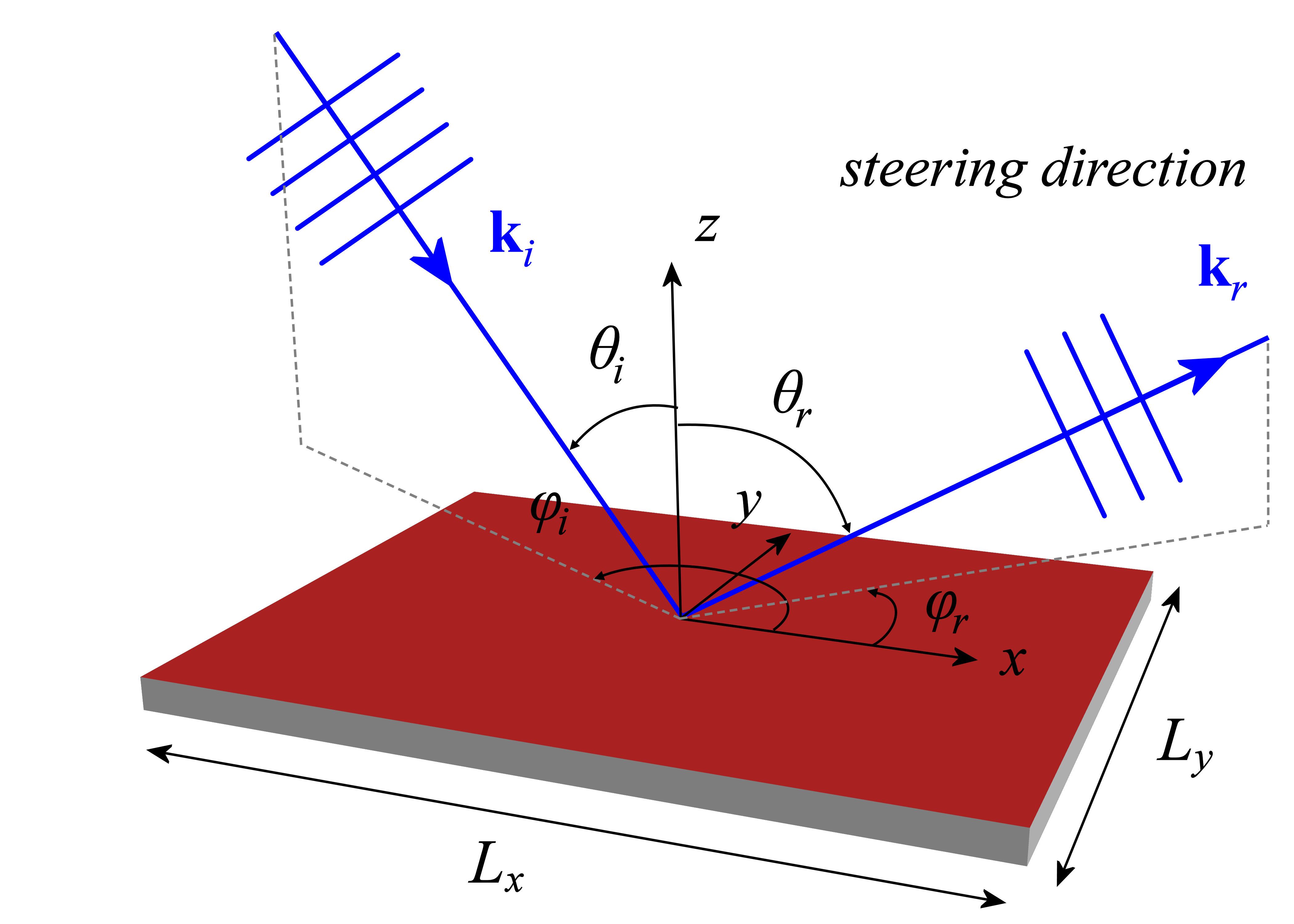}
	\caption{RIS as radiating sheet. The RIS is a continuous surface of size $L_x \times L_y$, which steers incident plane waves. The wavevectors $\textbf{k}_i$ and $\textbf{k}_r$ correspond to the incident ($i$) and reflected ($r$) wave and are characterized by the elevation and azimuth angles $\theta_i, \phi_i$ and $\theta_r, \phi_r$ respectively. Here we consider waves traveling on the $xz$-plane, i.e. $\phi_i=\pi$ and $\phi_r=0$.}
    	\label{fig:fig03}
\end{figure}
%
%
%
\begin{align}
    P_r = A_r \frac{k_0^2}{2Z_0} \left(\Theta_r \left|I_r\right|^2 + \Theta_i \left|I_i\right|^2 + \Theta_{ir} 2 Re(I_i I_r^*)\right),
    \label{Eq:EqM5}
\end{align}
where
\begin{align}
    \begin{split}
        \Theta_r(\theta,\phi) = \sin^2{\phi}(1+\cos{\theta}\cos{\theta_r})^2\\
        +\cos^2{\phi}(\cos{\theta}+\cos{\theta_r})^2,
    \end{split}
    \label{Eq:EqM6}
\end{align}
\begin{align}
    \begin{split}
        \Theta_i(\theta,\phi) = \sin^2{\phi}(1-\cos{\theta}\cos{\theta_i})^2\\
        +\cos^2{\phi}(\cos{\theta}-\cos{\theta_i})^2,
    \end{split}
    \label{Eq:EqM7}
\end{align}
\begin{align}
    \begin{split}
        \Theta_{ir}(\theta,\phi) = \sin^2{\phi}(1+\cos{\theta}\cos{\theta_r})(1-\cos{\theta}\cos{\theta_i})\\
+\cos^2{\phi}(\cos{\theta}-\cos{\theta_i})(\cos{\theta}+\cos{\theta_r}),
    \end{split}
    \label{Eq:EqM8}
\end{align}
with $\theta,\phi$ denoting the elevation and azimuth angles, respectively, of the far-field observation point. $I_r, I_i$ are integrals involving the magnitudes of the local (reflected, incident) fields on the RIS surface and are given in Appendix~\ref{Sec:Appendix_B}. Note that, to derive Eq.\@(\ref{Eq:EqM5}), we made no assumptions on the amplitude and phase of the incident wave and, therefore, the received power can be calculated for any illumination conditions. For plane wave illumination, $I_r, I_i$ can be expressed analytically and, as we show in Appendix~\ref{Sec:Appendix_C}, for sufficiently large RIS, the terms $\Theta_i \left|I_i\right|^2, \Theta_{ir} 2 Re(I_i I_r^*)$ are only a small correction to the leading term $\Theta_r \left|I_r\right|^2$ in Eq.\@(\ref{Eq:EqM5}). Hence, the received power is simply $P_r = A_r (k_0^2/2Z_0) \Theta_r \left|I_r\right|^2$, or:
\begin{multline}
    P_r = A_r \frac{k_0^2}{2Z_0} \Theta_r |\Gamma_0 E_{i}|^2 \left(\frac{L_xL_y}{4\pi r}\right)^2 \times \\ \left|
   \frac{\sin{\left(\frac{k_0L_x}{2}\left(\sin{\theta}\cos{\phi}-\sin{\theta_r}\right)\right)}}{\frac{k_0L_x}{2}\left(\sin{\theta}\cos{\phi}-\sin{\theta_r}\right)} \times\frac{\sin{\left(\frac{k_0L_y}{2}\sin{\theta}\sin{\phi}\right)}}{\frac{k_0L_y}{2}\sin{\theta}\sin{\phi}}\right|^2. 
    \label{Eq:EqM9}   
\end{multline}
Interestingly, the result of Eq.\@(\ref{Eq:EqM9}) can be derived directly by omitting the incident field in the boundary conditions and using only the reflected field (see Appendix~\ref{Sec:Appendix_B}). This is equivalent to treating the RIS as a radiating aperture \cite{OrfanidisBOOK}, as has been adopted in \cite{Tretyakov2021, Stratidakis2021a, Stratidakis2021b, Stratidakis2022}.
\section{Relation between models}
\begin{figure}[t!]
\centering
		\includegraphics[width=1\linewidth]{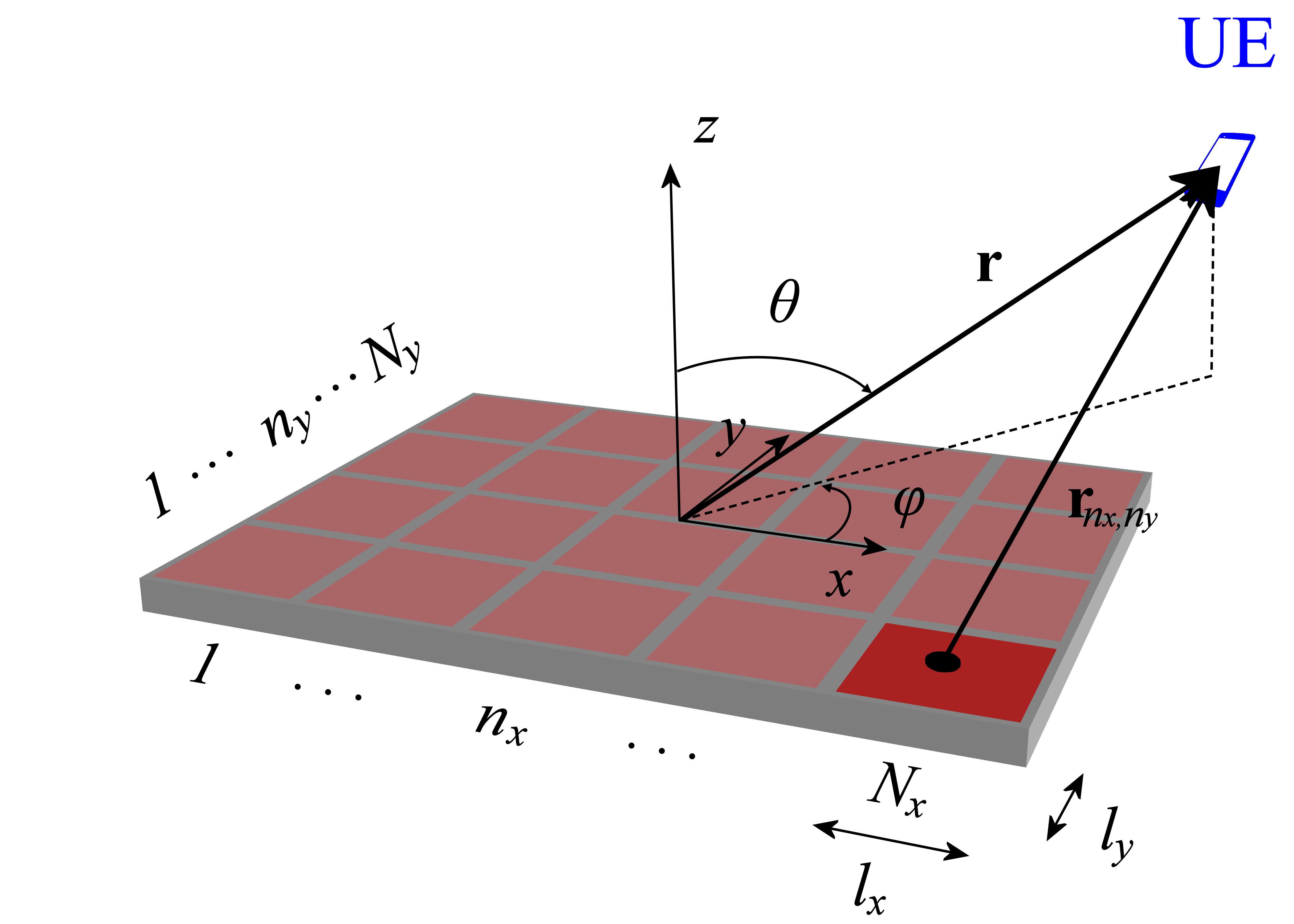}
	\caption{Connection between MIMO and sheet model. The continuous radiating sheet of size $L_x \times L_y$ is replaced by $N_x \times N_y$ dipoles, which are located at the center of each individual small rectangle of size $l_x \times l_y$. The total field at the UE is expressed as a discrete sum of all contributions from each dipole. Vectors $\textbf{r}$ and $\textbf{r}_{n_x,n_y}$ are the position vectors of the UE in the global coordinate system of the RIS and the local coordinate system of each dipole, respectively.}
    	\label{fig:fig04}
\end{figure}

\noindent To find a correspondence between the MIMO and sheet parameters, we need to spatially quantize the continuous properties of the sheet, and we therefore divide the latter into $N_y$ rows and $N_x$ columns. As a result, the surface of total area $L_x\times L_y$ is replaced by a grid of $N_x\times N_y$ dipoles, each placed at the center of a small rectangle of area of $l_x\times l_y$, and the integrals involved in the calculations of the previous section are discretized accordingly. For small enough periodicity, we can consider elements inducing only electric and/or magnetic dipoles and neglect the higher-order multipoles, which is reasonable for RIS element sizes in the order of $\approx \lambda_0/5$ or less. Following the steps outlined in Appendix~\ref{Sec:Appendix_D} we find that the electric field of the ($n_x,n_y$) dipole at the observation point $\textbf{r}$ is expressed as:
\begin{multline}
    \textbf{E}_r^{n_x,n_y} = \frac{1}{4\pi \epsilon_0} \frac{e^{-j k_0 r_{n_x,n_y}}}{r_{n_x,n_y}}k_0^2 
    [ (\hat{\textbf{r}}_{n_x,n_y}\times \textbf{p}_{n_x,n_y})\times \hat{\textbf{r}}_{n_x,n_y} \\ -(\hat{\textbf{r}}_{n_x,n_y}\times \frac{\textbf{m}_{n_x,n_y}}{Z_0}) ],
   \label{Eq:EqM10}
\end{multline}
where $\textbf{p}_{n_x,n_y}, \textbf{m}_{n_x,n_y}$ are the electric and magnetic dipole moments of the ($n_x,n_y$) element, respectively, $r_{n_x,n_y} = |\textbf{r}_{n_x,n_y}|$ and $\hat{\textbf{r}}_{n_x,n_y} = \textbf{r}_{n_x,n_y}/r_{n_x,n_y}$ is the unit vector, as illustrated in Fig.\,\ref{fig:fig04}. 
%
%
%
The dipole moments express the connection between the fields and the discretized electric and magnetic currents, however provide no knowledge about the actual properties of the RIS elements. To introduce the metasurface properties we may express $\textbf{p}_{n_x,n_y}, \textbf{m}_{n_x,n_y}$ as functions of the RIS element polarizabilities and the incident fields, i.e. as  \cite{TretyakovBOOK, SurfaceEMBOOK}:
\begin{subequations}
\begin{gather}
        \textbf{p}_{n_x,n_y} = \hat{\alpha}_{EE}\textbf{E}_i^{n_x,n_y} + \hat{\alpha}_{EM}\textbf{H}_i^{n_x,n_y}, \\
        \textbf{m}_{n_x,n_y} = \hat{\alpha}_{ME}\textbf{E}_i^{n_x,n_y} + \hat{\alpha}_{MM}\textbf{H}_i^{n_x,n_y},
\end{gather}
   \label{Eq:EqM11}
\end{subequations}
where $\hat{\alpha}_{EE}, \hat{\alpha}_{MM}, \hat{\alpha}_{EM}$, and $\hat{\alpha}_{ME}$ are the electric, magnetic, electromagnetic, and magnetoelectric (responsible for bi-anisotropy) polarizabilities, respectively, of an individual element, which are tensorial in their most general form. It is important to note that for a reflecting surface that does not allow transmission, both electric and magnetic moments are necessary \cite{Asadchy2016}. To steer TE waves on the $xz$-plane, we may use $\textbf{p}_{n_x,n_y} = p_{n_x,n_y} \hat{\textbf{y}}, \textbf{m}_{n_x,n_y} = m_{n_x,n_y} \hat{\textbf{x}}$ and, for simplicity, consider a non-bianisotropic RIS ($\hat{\alpha}_{EM} = \hat{\alpha}_{ME} = 0$), with polarizabilities $\hat{\alpha}_{EE}, \hat{\alpha}_{MM}$. A RIS with such general properties leads to a scattered field that, at distance $r_{n_x,n_y}$, has amplitude:
\begin{align}
    E_r^{n_x,n_y} = \frac{1}{4\pi \epsilon_0} \frac{e^{-j k_0 r_{n_x,n_y}}}{r_{n_x,n_y}}k_0^2 \alpha_{n_x,n_y}^{sheet}E_i^{n_x,n_y},
    \label{Eq:EqM12}
\end{align}
where
\begin{multline}
    \alpha_{n_x,n_y}^{sheet} = \Biggr[\cos^2{\phi}\left(\hat{\alpha}_{EE}-\hat{\alpha}_{MM}\frac{\cos{\theta}\cos{\theta_i}}{Z_0}\right)^2 \\ +\sin^2{\phi}\left(\hat{\alpha}_{EE}-\hat{\alpha}_{MM}\frac{\cos{\theta}\cos{\theta_i}}{Z_0}\right)^2 \Biggr]^{1/2}
    \label{Eq:EqM13}
\end{multline}
is the dipole polarizability of the ($n_x,n_y$) RIS element. At this point, we have made no assumptions about the values of the polarizabilities, which can be introduced from simulations and experiments. Here, we will use the general properties that were analytically derived in \cite{Asadchy2016}, are consistent with Maxwell's equations for reflecting sheets, and can be written in the form:
\begin{subequations}
\begin{gather}
        \hat{\alpha}_{EE} = \frac{l_x l_y}{j \omega} \left(\frac{\cos{\theta_i}}{Z_0}-\Gamma_{n_x,n_y}\frac{\cos{\theta_r}}{Z_0}\right), \\
        \hat{\alpha}_{MM} = \frac{l_x l_y}{j \omega} \frac{Z_0}{\cos{\theta_i}} \left(1+\Gamma_{n_x,n_y}\right),
\end{gather}
   \label{Eq:EqM14}
\end{subequations}
where $\Gamma_{n_x,n_y}$ is the sheet reflection coefficient calculated at the position of the $(n_x,n_y)$ element. Substitution of Eq.\@(\ref{Eq:EqM14}) in Eq.\@(\ref{Eq:EqM13}), yields the sheet polarizability:
\begin{align}
    \alpha_{n_x,n_y}^{sheet} = l_x l_y\frac{j \epsilon_0}{k_0}\sqrt{\Gamma^2_{n_x,n_y} \Theta_r + \Theta_i + 2 \Gamma_{n_x,n_y} \Theta_{ir}}.
    \label{Eq:EqM15}
\end{align}
For sufficiently large RIS (see Appendix C) we may again treat the RIS as a radiating aperture and neglect the incident field when applying the boundary conditions to derive Eq.\@(\ref{Eq:EqM14}). In this case the terms related to $\Theta_i, \Theta_{ir}$ vanish and the polarizability takes the simpler form:
\begin{align}
    \alpha_{n_x,n_y}^{sheet} = l_x l_y\frac{j \epsilon_0}{k_0} \Gamma_{n_x,n_y} \sqrt{\Theta_r(\theta, \phi)}.
    \label{Eq:EqM16}
\end{align}
Next, to find an equivalence between the MIMO and sheet model, we will require that the discrete scatterers of both approaches are characterized by the same polarizabilities, i.e. $\alpha^{MIMO}_{n_x,n_y} = \alpha^{sheet}_{n_x,n_y}$. For $\alpha^{sheet}_{n_x,n_y}$ we will use Eq.\@(\ref{Eq:EqM16}) that was derived specifically for elements characterized by the properties given by Eq.\@(\ref{Eq:EqM14}), and our conclusions can be directly extended to any other combination of electric and magnetic properties, using the more general Eq.\@(\ref{Eq:EqM13}).
\subsection{Correspondence between MIMO and sheet parameters}
\noindent  Often, it is assumed that $R_{n_x,n_y} \equiv \Gamma_{n_x,n_y}$ and $A_{UC}\equiv l_xl_y$ \cite{DiRenzo2021, Alexiou2021}. As a result, by requiring $\alpha^{MIMO}_{n_x,n_y} = \alpha^{sheet}_{n_x,n_y}$, the remaining properties of the MIMO elements can be associated with the properties of the radiating dipoles of the discretized sheet model as:
\begin{subequations}
\begin{gather}
    R_{n_x,n_y} \equiv \Gamma_{n_x,n_y} \\
    A_{UC} \equiv l_x l_y, \\
    G_{UC} = \frac{4\pi}{\lambda_0^2} A_{UC} \equiv \frac{4\pi}{\lambda_0^2} l_x l_y, \\
    U_{UC}^{AP} = 1, \\    
    U_{UC}^{UE} \equiv \frac{\Theta_r(\theta,\phi)}{4},
\end{gather}
    \label{Eq:EqM17}
\end{subequations}
where the angles $\theta, \phi$ that are associated with the observation point in the sheet model, correspond to the UE location in the MIMO model. With Eqs.\@(\ref{Eq:EqM17}), Eqs.\@(\ref{Eq:EqM3}) and \@(\ref{Eq:EqM16}) become identical, ensuring the equivalence between the two models. \\
%
%
%
\begin{figure}[t!]
\centering
		\includegraphics[width=1\linewidth]{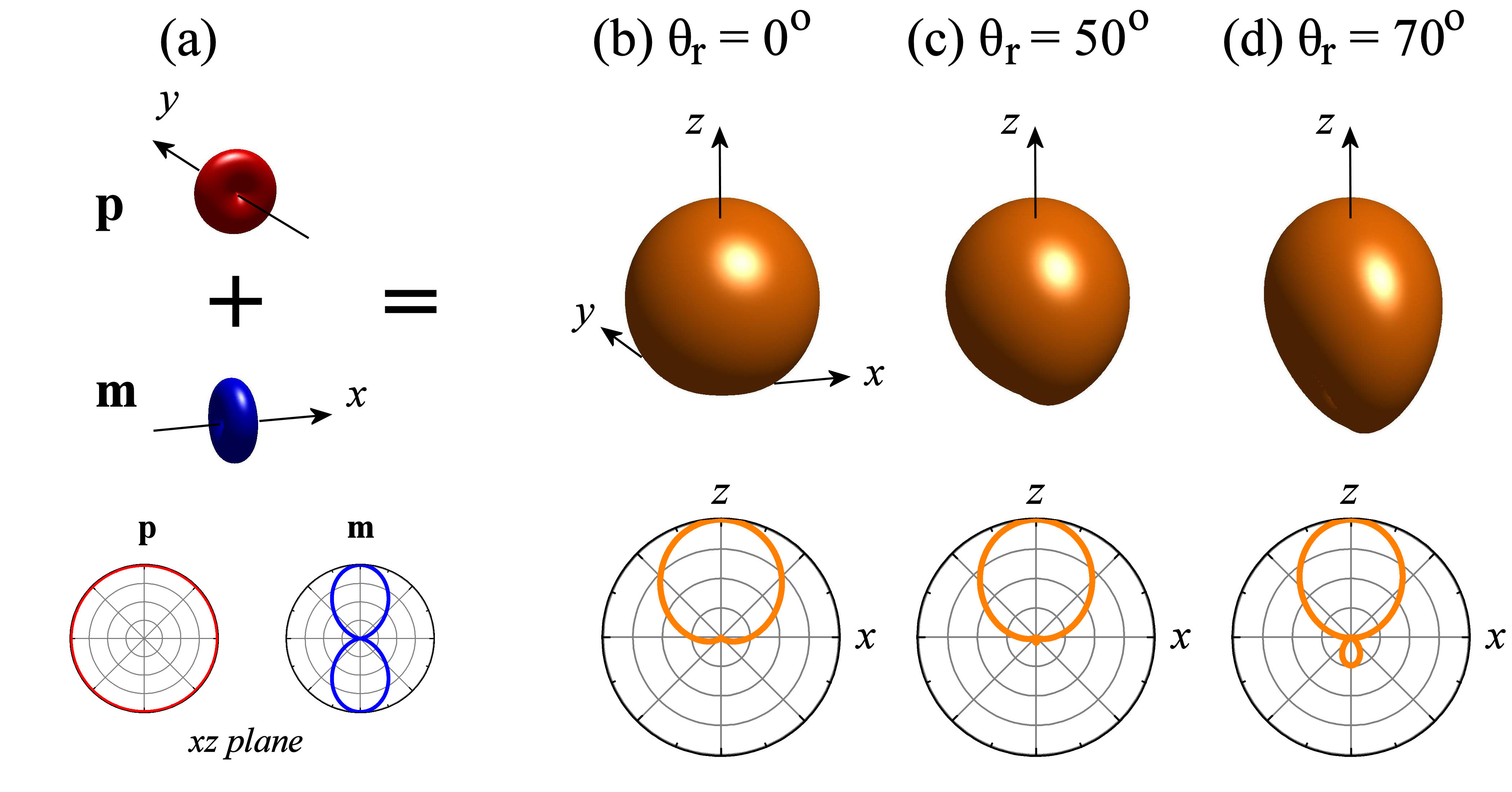}
	\caption{Radiation pattern $U_{UC}^{UE}$ of a single RIS element under the sheet-MIMO equivalence given by Eq.\@(\ref{Eq:EqM17}). The total radiation pattern is a weighted mixture of an electric ($\textbf{p}$) and a magnetic ($\textbf{m}$) dipole, as shown in (a). Examples for a RIS reflecting towards (b) $\theta_r=0^o$, (c) $\theta_r=50^o$, and (d) $\theta_r=70^o$. The shape of the radiation pattern depends strongly on the chosen steering angle, as a result of the different phase and amplitude balance between the $\textbf{p}$ and $\textbf{m}$ constituents.}
    	\label{fig:fig06}
\end{figure}
%
%
\indent Evidently, the derived properties in Eq.\@(\ref{Eq:EqM17}) can hardly represent scattering elements with fixed radiation pattern. The reason is that, because both electric and magnetic moments are necessary for a  non-penetrable surface, in an attempt to find the effective MIMO polarizability we are trying to fit a variable mixture of electric and magnetic properties into a well-defined radiation pattern, which is not possible.
The radiation pattern of the ($n_x,n_y$) element necessarily derives from a weighted superposition of $\textbf{p}_{n_x,n_y}$ and $\textbf{m}_{n_x,n_y}$, as expressed by Eq.\@(\ref{Eq:EqM10}) and illustrated in Fig.\,\ref{fig:fig06}(a). As the desired reflection angle changes, the relative phase and amplitude between $\textbf{p}_{n_x,n_y}$ and $\textbf{m}_{n_x,n_y}$ changes, leading to a radiation pattern with shape that depends on the chosen reflection angle, as shown here for $\theta_r=0^o$ [Fig.\,\ref{fig:fig06}(b)], $\theta_r=50^o$ [Fig.\,\ref{fig:fig06}(c)] and $\theta_r=70^o$ [Fig.\,\ref{fig:fig06}(d)]. Although this could be a possibility, what is here sought after is scatterers with radiation pattern that (a) has the same shape for both reception and emission operation, and (b) does not depend on the chosen reflection angle; wave steering should result from the different excitation phase among the RIS elements, all having identical radiation patterns.
\subsection{Correspondence consistent with antenna theory}
\noindent In antenna theory \cite{BalanisBOOK}, it is convenient to express all relevant quantities with respect to the radiation intensity $U(\theta,\phi)=r^2S_{rad}(\theta,\phi)$ (W/sr), where $S_{rad}$ is the power density emitted by the antenna at distance $r$; the total power is obtained by integrating over the entire solid angle $P_{rad}=\int_0^{2\pi}\int_0^{\pi}U(\theta,\phi)\sin{\theta}d\theta d\phi$. \\
\indent Under the MIMO approach, we can attribute to each RIS element a radiation pattern $U(\theta,\phi) = U_{max}U_0(\theta,\phi)$, where $U_{max}$ is the maximum value of $U(\theta,\phi)$ and $U_0(\theta,\phi)$ is the normalized radiation pattern. Then, the antenna parameters of the RIS elements, including the directivity $D_{UC}$, gain $G_{UC}$ and receiving aperture $A_{UC}$, are expressed in terms of $U(\theta,\phi)$ as:
\begin{subequations}
    \begin{gather}
    U_{UC}^{AP} \equiv U_0(\theta^{AP}_{n_x,n_y},\phi^{AP}_{n_x,n_y}), \\ 
    U_{UC}^{UE} \equiv U_0(\theta^{UE}_{n_x,n_y},\phi^{UE}_{n_x,n_y}), \\     
    D_{UC} = 4\pi \frac{U_{max}}{\int_0^{2\pi} \int_0^{\pi} U(\theta, \phi) \rm{sin}\theta d\theta d\phi} \\
    G_{UC} = e_0 D_{UC}  \\
    A_{UC} = \frac{\lambda_0^2}{4\pi} G_{UC},
    \end{gather}
    \label{Eq:EqM18}
\end{subequations}
where $e_0$ accounts for the antenna efficiency, which models possible absorption and scattering losses ($0<e_0<1$), and $\theta_{n_x,n_y}, \phi_{n_x,n_y}$ denote the elevation and azimuth angle from the ($n_x,n_y$) element to the transmitter (AP) and the receiver (UE), as previously illustrated in Fig.\,\ref{fig:fig02}. Note that, under plane wave illumination, all elements are excited under the same angle of incidence, i.e. $\theta^{AP}_{n_x,n_y} \equiv \theta_i$, $\phi^{AP}_{n_x,n_y} \equiv \phi_i$. Hence, $U_{UC}^{AP}$ is a common constant for all elements and we can simply write $U_{UC}^{AP} \equiv U_0(\theta_i, \phi_i)$. Additionally, in the far-field of the RIS where $\theta^{UE}_{n_x,n_y} \approx \theta$, $\phi^{UE}_{n_x,n_y} \approx \phi$ we may simplify $U_{UC}^{UE} \equiv U_0(\theta, \phi)$, as already adopted in Eq.\@(\ref{Eq:EqM17}e). \\
\indent In the above analysis, we have defined all relevant antenna parameters except for $R_{n_x,n_y}$, which is associated with the scattering strength of the RIS and is, therefore, closely related to $e_0$. At this point we are free to set $R_{n_x,n_y}=\Gamma_{n_x,n_y}$ and use $e_0$ as a separate parameter or we may incoroporate $e_0$ into a new reflection coefficient $R_{n_x,n_y}=e_0\Gamma_{n_x,n_y}$ and use the directivity instead of the gain in Eq.\@(\ref{Eq:EqM18}). Alternatively, we may define a new antenna efficiency or correction factor $e_0'=e_0 \Gamma_{n_x,n_y}$ and omit $R_{n_x,n_y}$ from the analysis; all approaches are mathematically equivalent. Requiring $\alpha^{MIMO}_{n_x,n_y} = \alpha^{sheet}_{n_x,n_y}$, we find using Eq.\@(\ref{Eq:EqM3}) and Eq.\@(\ref{Eq:EqM16}) that the unknown $e_0$ must satisfy:
\begin{multline}
    e_0 \equiv \frac{k_0^3}{2\pi \epsilon_0} \frac{\alpha^{sheet}_{n_x,n_y}}{R_{n_x,n_y} D_{UC} \sqrt{U_{UC}^{AP} U_{UC}^{UE}}} = \\ = \frac{4\pi}{D_{UC}}\frac{l_xl_y}{\lambda_0^2} \sqrt{\frac{\Theta_r(\theta,\phi)}{4U_{UC}^{AP} U_{UC}^{UE}}}
    \label{Eq:EqM19}
\end{multline}
\noindent A RIS described by antenna elements with properties given by Eqs.\@(\ref{Eq:EqM18}),\@(\ref{Eq:EqM19}) is consistent with the antenna theory and reproduces successfully the scattered field predicted by the sheet model. Note that, due to $\Theta_r$ and $U_{UC}^{UE}$, the correction factor $e_0$ that adjusts the maximum intensity is a function of the observation angle; this implies that $e_0$, besides adjusting the gain, also modifies the shape of the radiation pattern. \\
\indent To gain some insight into the correction factor $e_0$, next we will consider some typical antenna radiation patterns that are commonly used in related works to describe the response of subwavelength scatterers.
%
%
%
%
\begin{figure}[t!]
\centering
		\includegraphics[width=1\linewidth]{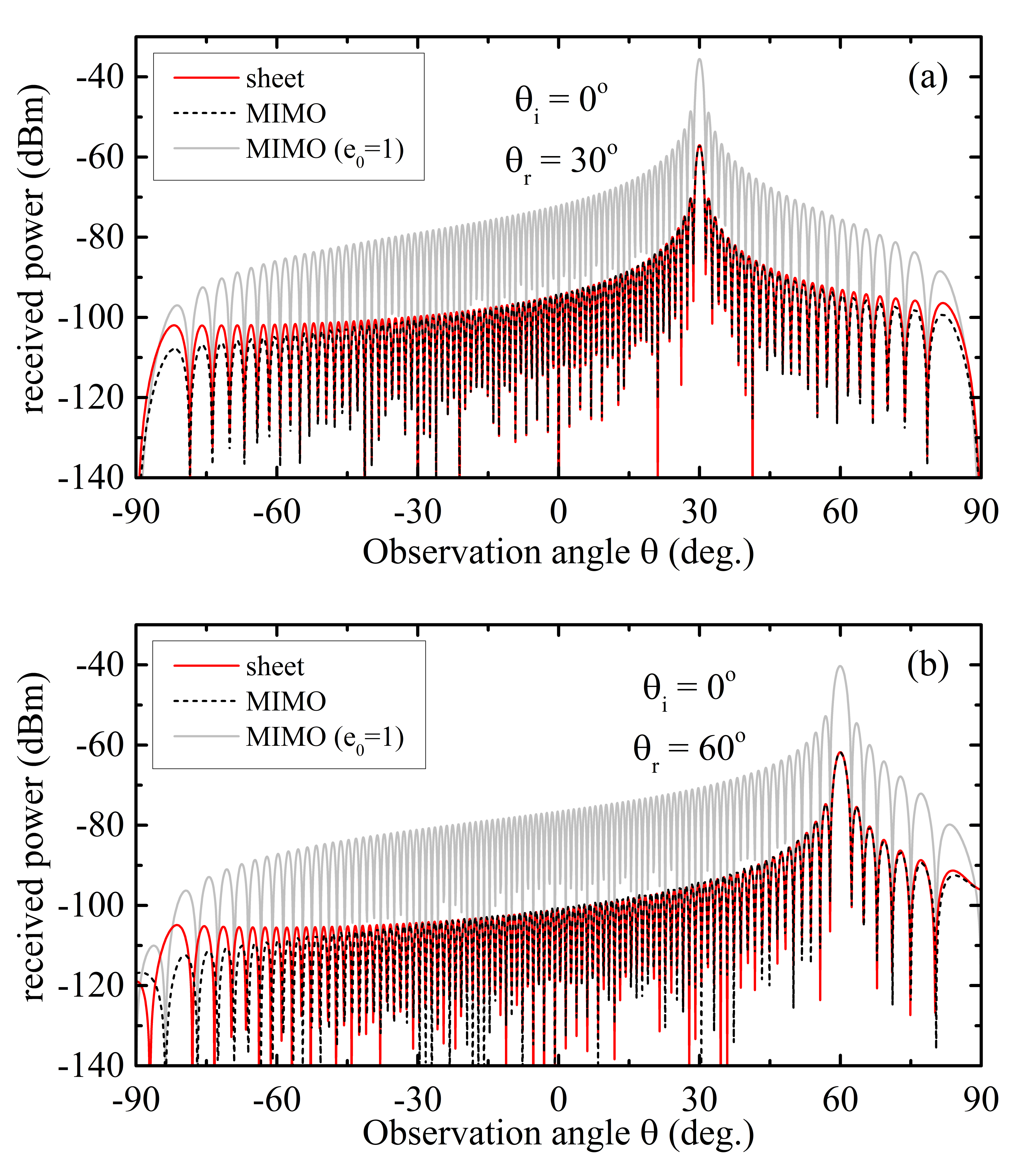}
	\caption{Equivalence between the sheet and MIMO model for RIS elements modeled as antennas with radiation pattern $U_0(\theta,\phi)=\cos^2{\theta}$ and properties consistent with antenna theory. A TE-polarized wave with $E_i = 1$ V/m, normally incident on the RIS ($\theta_i = 0$), is steered by the RIS with $|\Gamma_0|=1$ towards (a) $\theta_r = 30^o$, (b) $\theta_r = 60^o$. Without the correction factor, i.e. with $e_0 = 1$ (solid gray line), the antenna model overestimates the analytically predicted power (solid red line); the sheet-MIMO equivalence is restored upon incorporating $e_0$ (dashed black line). The RIS is operating at 150 GHz and consists of $250\times 250$ elements with periodicity $\lambda_0/5$ along both $x$ and $y$ directions, forming a reflecting surface of total size 10 cm $\times$ 10 cm. The received power is calculated at distance 20 m from the RIS, as a function of the observation angle $\theta$ that marks the UE position.}
    	\label{fig:fig07}
\end{figure}
%
%
\subsection{Examples of RIS element properties consistent with antenna theory}
\noindent The radiation intensity of the major lobe of many antennas is frequently represented by $U_0(\theta,\phi) = \cos^q{\theta}$, where the positive real $q$ modifies the antenna directivity \cite{BalanisBOOK}. Using this form for $U_0$, the properties of the RIS elements are analytically expressed as:
\begin{subequations}
\begin{gather}
    U_{UC}^{AP} = \cos^q{\theta_i}, \\    
    U_{UC}^{UE} = \cos^q{\theta}, \\
    D_{UC} = 2(q+1), \\
    G_{UC} = 2(q+1)e_0(\theta,\phi), \\
    A_{UC} = \frac{\lambda_0^2}{4\pi}2(q+1)e_0(\theta,\phi), \\
    e_0(\theta,\phi) = \frac{4\pi}{2(q+1)}\frac{l_x l_y}{\lambda_0^2}\sqrt{\frac{\Theta_r(\theta,\phi)}{4\cos^q{\theta_i}\cos^q{\theta}}}.
\end{gather}
    \label{Eq:EqM20}
\end{subequations}
\indent As an example, let us consider a RIS of size 10 cm $\times$ 10 cm, operating at 150 GHz ($\lambda_0=2$ mm), consisting of $250\times 250$ elements, with periodicity $\lambda_0/5$ along both $x$ and $y$ directions and radiation pattern $U_0(\theta,\phi)=\cos^2{\theta}$ ($q=2$). A $y$-polarized plane wave propagating on the $xz$-plane with $E_i = 1$ V/m illuminates the RIS at normal incidence ($\theta_i=0$) and the RIS redirects the wave towards angle $\theta_r$ with $|\Gamma_0|=1$. The receiver can move freely on the same plane at constant distance $r=20$ m from the RIS and has antenna gain $G_r =$ 20 dB. The received power, as a function of the observation angle $\theta$ that marks the UE position, is shown in Fig.\,\ref{fig:fig07}(a) for $\theta_r=30^o$ and in Fig.\,\ref{fig:fig07}(b) for $\theta_r=60^o$. In both examples the solid red line is the plot of Eq.\@(\ref{Eq:EqM5}) for the sheet model, and the dashed black line is the numerically calculated Eq.\@(\ref{Eq:EqM4}) for the MIMO model, using the parameters given by Eq.\@(\ref{Eq:EqM20}). The numerically calculated received power without using the correction factor is also shown with the solid gray lines. Clearly, without the correction factor, i.e. by setting $e_0=1$, the MIMO model overestimates the power density practically across the entire observation range and, importantly, at the steering angle $\theta_r$, where the redirected power density is maximized.
The results demonstrate that both approaches lead to the same received power, except perhaps for very large observation angles, where the power is practically negligible and, hence, the discrepancy is irrelevant. In general, deviations result from two sources: (a) from the omitted terms $\Theta_i, \Theta_{ir}$ in Eq.\@(\ref{Eq:EqM16}) and (b) by replacing the continuous integrals in Eq.\@(\ref{Eq:EqB7}) with discrete sums. The former source of discrepancy can be lifted either with larger RIS (see Appendix~\ref{Sec:Appendix_C}) or with using the full polarizability of Eq.\@(\ref{Eq:EqM15}); the latter is lifted with smaller RIS elements. Note that we can always lift the latter source of discrepancy entirely, if we replace the global integrals in Eq.\@(\ref{Eq:EqB7}) with sums of integrals within each small rectangle and perform the individual integrations (see Appendix~\ref{Sec:Appendix_D}).
\subsection{Commonly used RIS element properties}
\noindent Last, we would like to note that in some works as \cite{DiRenzo2021, Alexiou2021}, the RIS element properties are considered as:
\begin{subequations}
    \begin{gather}
        U_{UC}^{AP} = \cos^q{\theta_i}, \\    
        U_{UC}^{UE} = \cos^q{\theta}, \\
        A_{UC} = l_x l_y, \\
        G_{UC} = 2(q+1) \\
        R_{n_x,n_y} = \Gamma_{n_x,n_y}.    
    \end{gather}
    \label{Eq:EqM21}
\end{subequations}
This model borrows elements from the antenna theory (form of $U_{UC}$ and corresponding $G_{UC}$), however considers a $100\%$ aperture efficiency ($A_{UC}/l_xl_y$) and does not account for the necessary correction factor. Therefore, the scattering predicted by this model is overestimated.
 \section{Analysis of correction factor}
\noindent The correction factor captures the discrepancy in the power predicted by the two approaches; note that the ratio of the received power calculated with the MIMO model over the received power calculated with the sheet model is simply the ratio of the squared magnitudes of the corresponding reflection coefficients, yielding $1/e_0^2$. For small incident angles where $\cos^q{\theta_i} \to 1$, and taking into account that $l_x,l_y \ll \lambda_0$ and $\Theta_r(\theta,\phi)/4 \leq$ 1, with a simple inspection of Eq.\@(\ref{Eq:EqM20}f), we see that $1/e_0^2>1$, i.e. the MIMO model overestimates the received power with respect to the sheet model. The opposite occurs for large incident angles where $\cos^q{\theta_i} \to 0$, however requires $q\gg 1$, i.e. very directive scatterers, which is hard to achieve with subwavelength elements. \\
\indent The presence of the factor $\Theta_r(\theta,\phi)$ in $e_0$ implies that, to achieve equivalence between the discrete and continuous models, the radiation pattern of the RIS elements must depend explicitly on the steering angle $\theta_r$. This means that, in realistic implementations where the scatterers have a fixed radiation pattern, deviations should be expected and, for the same incident wave, the magnitude of the scattered wave will depend on the shape of the scatterer radiation pattern. Importantly, due to the dependency of $e_0$ on $\theta_r$, the discrepancy in the received power between the discrete and continuous model is expected, in general, to change with the steering angle. \\
%
%
%
\begin{figure}[t!]
\centering
		\includegraphics[width=1\linewidth]{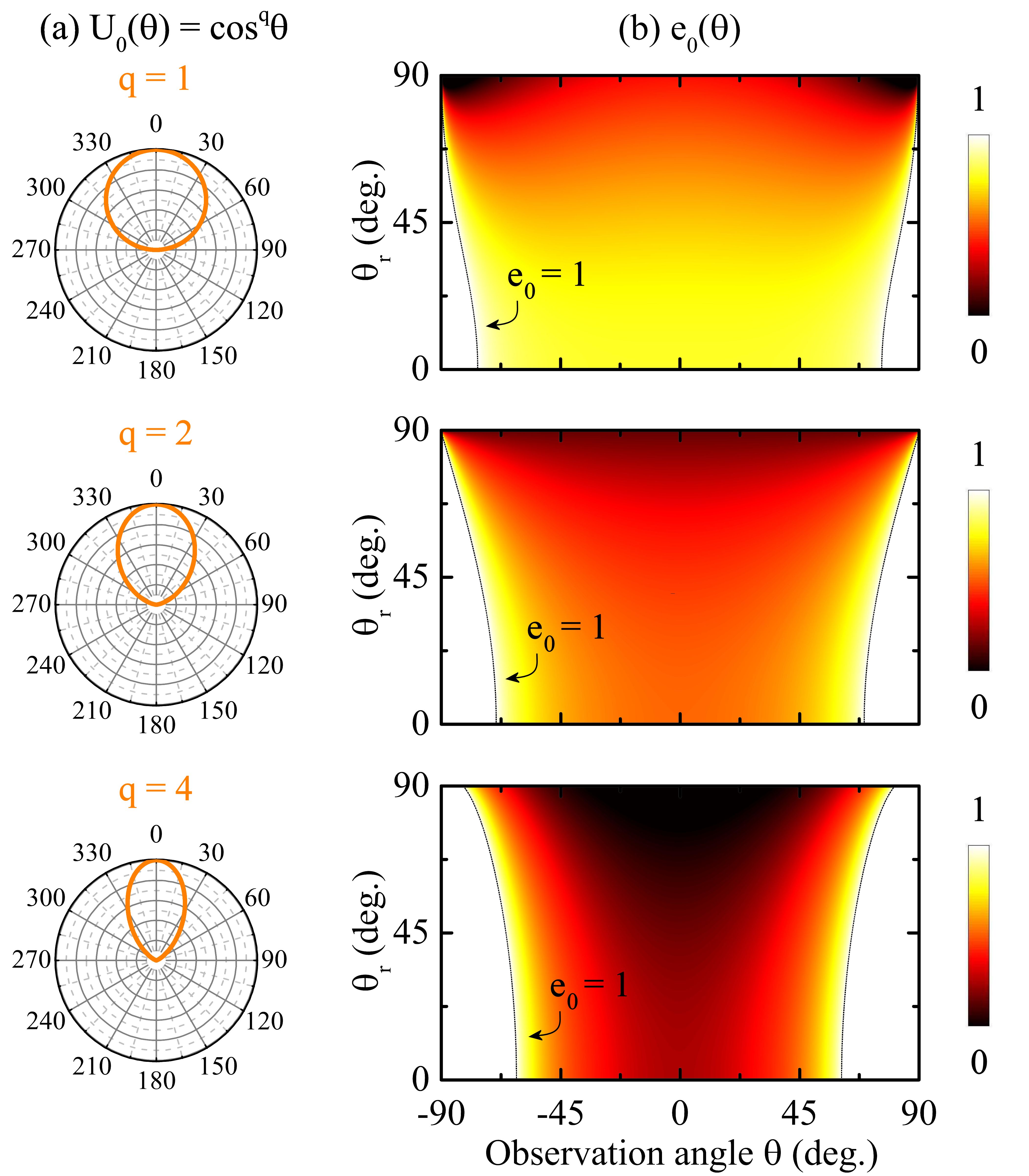}
	\caption{Correction factor $e_0$ for RIS elements modeled as antennas with radiation pattern $U(\theta,\phi)=\cos^q{\theta}$ and properties consistent with antenna theory. (a) Polar plot of RIS element radiation pattern for $q$ = 1, 2 and 4. (b) Correction factor on the $xz$-plane ($\phi=0$) for elements with $l_x = l_y=\lambda_0/2$, as a function of the steering angle $\theta_r$ for a normally incident wave ($\theta_i=0^o$). The colormap has been saturated within the range $(0 . . . 1)$ to emphasize the wide region within the dashed lines, where $e_0<1$ and the MIMO model overestimates the scattered field if the correction factor is not taken into account. With decreasing element size the region with $e_0<1$ expands towards $|\theta|\rightarrow 90$ deg., as $e_0$ balances the increasing density of scatterers, ensuring the same scattered field.}
    	\label{fig:fig08}
\end{figure}
%
\indent To gain more insight into the correction factor $e_0$, in Fig.\,\ref{fig:fig08}, we consider a RIS consisting of elements with tunable radiation pattern of the form $U_0(\theta,\phi)=\cos^q{\theta}$, i.e. complying with Eqs.\@(\ref{Eq:EqM20}). The radiation pattern for elements with $q=1,2,4$ is shown in Fig.\,\ref{fig:fig08}(a). For each case, in Fig.\,\ref{fig:fig08}(b), $e_0$ is calculated for elements with $l_x=l_y=\lambda_0/2$ as a function of the steering angle $\theta_r$, for normal incidence ($\theta_i=0$). The dashed lines mark the boundary $e_0=1$, i.e. the special case where both the MIMO and sheet provide the same scattered field. Note that $e_0<1$ for the major range of the observation angle, i.e. the MIMO model overestimates the scattered field if the correction factor is not taken into account. With decreasing element size, the density of the scatterers increases and $e_0$ becomes smaller in order to ensure the same scattered field, pushing the $e_0=1$ boundary towards $|\theta|\rightarrow 90$ deg. The element size chosen in these examples practically marks the boundary between common MIMO systems and RISs: the former usually involve antenna separations larger than $\lambda_0/2$, contrary to the latter. \\
\indent To further quantify the analysis with respect to the RIS element directivity, it is instructive to calculate $e_0$ along the steering angle ($\theta=\theta_r$, $\phi=0$), where the UE is expected to be located. In this case the correction factor takes the simple form:
\begin{equation}
    e_0 = \frac{4\pi}{2(q+1)}\frac{l_xl_y}{\lambda_0^2}(\cos{\theta_i})^{-\frac{q}{2}}(\cos{\theta_r})^{1-\frac{q}{2}}.
    \label{Eq:EqM22}
\end{equation}
In Fig.\,\ref{fig:fig09}(a) we plot Eq.\@(\ref{Eq:EqM22}) as a function of the exponent $q$, to demonstrate the impact of the RIS element directivity on the dependence of $e_0$ on $\theta_r$. The results are normalized with $l_xl_y/\lambda_0^2$ in order to provide a universal diagram for RIS elements with radiation pattern of the form $\cos^q{\theta}$, under normal incidence ($\theta_i=0^o$); for different incident angles the results scale with $\cos^{-\frac{q}{2}}{\theta_i}$. For highly directive elements that promote forward emission strongly (e.g. for $q=4$), $e_0$ increases with increasing angle to compensate the preferential emission. On the contrary, for weakly directive elements that promote forward emission slightly (e.g. for $q=1$), $e_0$ decreases with increasing angle. The turning point between the two extremes occurs for elements with $q=2$, for which $\Theta_r = 4\cos^2{\theta_r}$ and, therefore, $U_{UC}^{UE}=\cos^2{\theta_r}$ exactly compensates $\Theta_r$ for all angles [see Eq.\@(\ref{Eq:EqM20})]. In this case the incident wave can be steered to any direction with constant $e_0$, i.e. the discrepancy between the sheet and MIMO model does not depend on $\theta_r$; this can be seen in Eq.\@(\ref{Eq:EqM22}) by setting $q=2$. To emphasize the qualitative change of $e_0$ across the critical value $q=2$ (marked with the dashed line), the colormap is saturated within the range $(1 \ldots 3$). \\
%
%
\begin{figure}[t!]
\centering
		\includegraphics[width=1\linewidth]{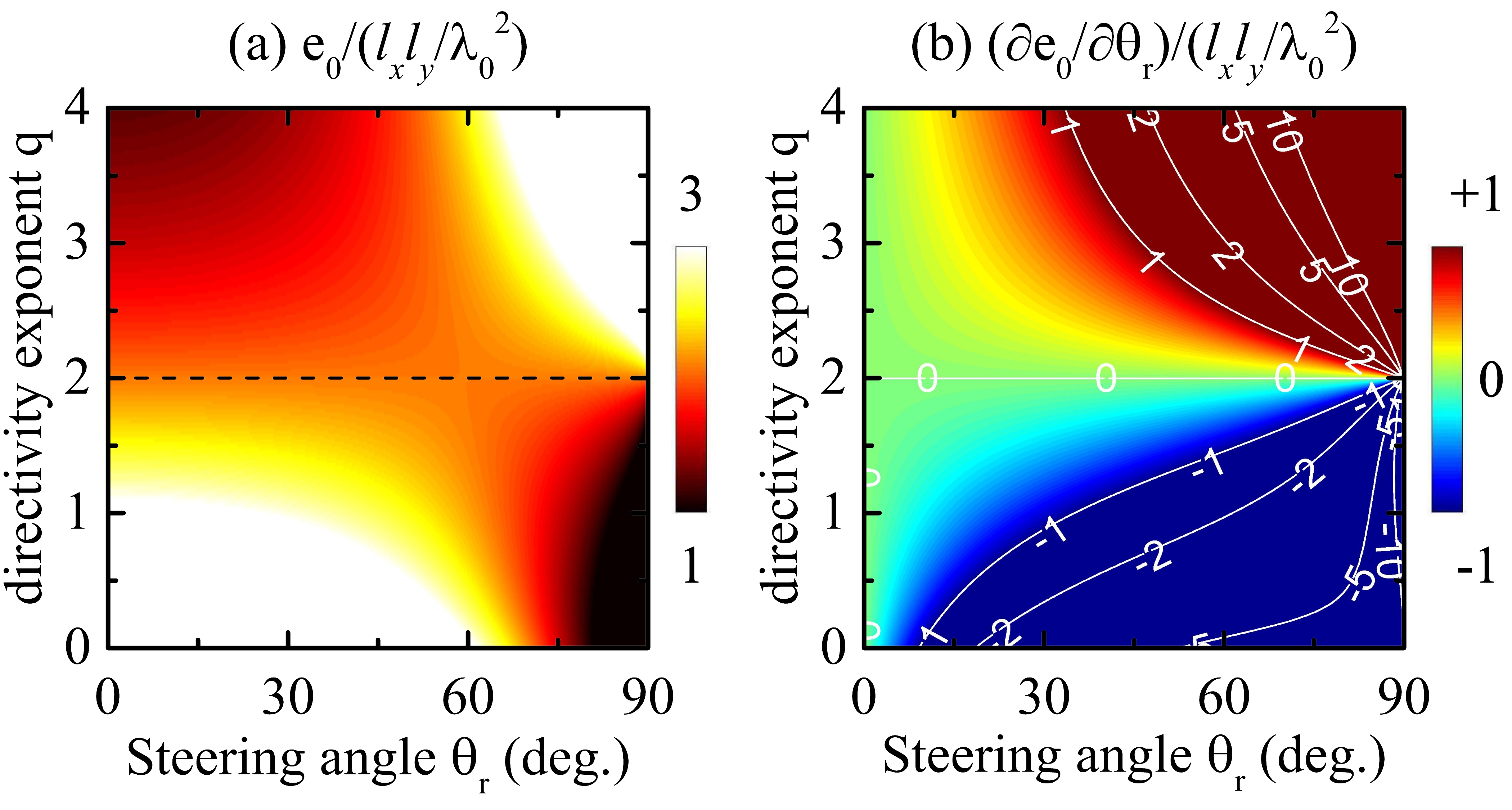}
	\caption{Sensitivity analysis of correction factor $e_0$ for elements with radiation pattern of the form $U_0(\theta,\phi)=\cos^q{\theta}$. Normalized plots of (a) $e_0$, and (b) $\frac{\partial e_0}{\partial\theta_r}$, as a function of the exponent $q$, which expresses the RIS element directivity.}
    	\label{fig:fig09}
\end{figure}
%
\indent To estimate the sensitivity of $e_0$ on the steering angle $\theta_r$ in Fig.\,\ref{fig:fig09}(b) we plot the derivative:
\begin{equation}
     \frac{\partial e_0}{\partial\theta_r} = (q-2)\frac{\pi}{q+1}\frac{l_x l_y}{\lambda_0^2}(\cos{\theta_i})^{-\frac{q}{2}}(\cos{\theta_r})^{-\frac{q}{2}}\sin{\theta_r},
     \label{Eq:EqM23}
\end{equation}
normalized with $l_xl_y/\lambda_0^2$. The results demonstrate that, for RIS elements with $q\sim 2$, the sensitivity of $e_0$ on $\theta_r$ practically remains weak for any desired reflection angle. The same can be achieved for RIS elements with arbitrary directivity, as long as the reflection angles are small. In essence, while the MIMO and sheet models are equivalent in the entire parameter space, there is a range [green region in Fig.\,\ref{fig:fig09}(b)] within which the sensitivity of $e_0$ on $\theta_r$ can be minimized for RIS elements with fixed radiation pattern. Note that, with decreasing RIS element size, the factor $l_x l_y/\lambda_0^2$ reduces and the sensitivity decreases, in turn expanding the respective range.
 \section{Impact of element coupling}
\noindent Typically, in MIMO systems, the antenna separation is usually large enough so that the coupling between antennas is negligible. In RISs, though, the elements are by design closely spaced and, depending on their particular implementation, the individual response of each RIS element could be possibly modified by the neighboring elements, in which case coupling cannot be neglected. \\
\indent Coupling can be always introduced implicitly in the properties of the RIS elements. For example, when a RIS is characterized by means of global parameters, such as reflectance and transmittance, the retrieved polarizabilities already contain the interactions and are, hence, collective properties. Theoretically, this corresponds to expressing the RIS element polarizabilities in terms of the incident fields only, instead of the sum of the incident field and the scattered fields \cite{TretyakovBOOK, Niemi2003, SurfaceEMBOOK}, and this is what we have considered so far [(see Eq.\@(\ref{Eq:EqM11})]. \\
\indent When the properties of individual elements do not include their interactions, e.g. as occurs when simulating isolated scatterers, then to characterize the global response of the RIS it is necessary to introduce coupling explicitly. Returning to the general picture of Eq.\@(\ref{Eq:EqM1}), the dipole moment of the $u^{th}$ element [for simplicity denoting the index pair ($n_x,n_y$)] is expressed for uncoupled elements as $\textbf{p}_u = \alpha_u^{MIMO} \textbf{E}^{AP}_u$, where $\alpha_u^{MIMO}$ is the element polarizability and $\textbf{E}^{AP}_u$ is the electric field at position $u$ due to the incident wave from the AP. When the interaction with the neighboring elements cannot be neglected, the dipole moment is written as:
\begin{equation}
    \textbf{p}_u = \alpha_u^{MIMO}  \left(\textbf{E}^{AP}_u - \sum_{v \neq u} \textbf{K}_{uv} \textbf{p}_v\right),
    \label{Eq:EqM24}
\end{equation}
where $\textbf{K}_{uv} \textbf{p}_v$ is the contribution to the electric field at position $u$ due to the element at position $v$ \cite{Draine1994}. The elements of matrix $\textbf{K}_{uv}$ are the coupling constants $\kappa_{uv}$, that express the interaction between the $u^{th}$ and $v^{th}$ element. To solve this equation self-consistently in terms of the unknown element dipole moments $\textbf{p}_u$ it is possible to reduce it to a matrix equation $\textbf{C} \textbf{P} =\textbf{E}^{AP}$, where $\textbf{P}=[\textbf{p}_1 \textbf{p}_2 ... \textbf{p}_{N_x\times N_y}]^T$, $\textbf{E}^{AP}=[\textbf{E}_1^{AP} \textbf{E}_2^{AP} ... \textbf{E}_{N_x\times N_y}^{AP}]^T$ and $\textbf{C}=\textbf{C}_{self}+\textbf{C}_{mutual}$ is a 3($N_x\times N_y$)$\times$3($N_x\times N_y$) matrix, with elements $\textbf{C}_{self,uv} =(\alpha_u^{MIMO})^{-1}\delta_{uv}\textbf{I}$, and $\textbf{C}_{mutual,uv} = \kappa_{uv}(1-\delta_{uv})\textbf{I}$. In this notation we have used the $3\times 3$ identity matrix $\textbf{I}$ and $\delta_{uv}$, the Kronecker delta, to emphasize that $\textbf{C}_{self}$ fills the $3\times3$ block-diagonal elements of $\textbf{C}$, while $\textbf{C}_{mutual}$ fills the remaining elements. The solution of this linear system is achieved by calculating the inverse of matrix $\textbf{C}$, which is essentially the polarizability matrix that expresses the connection between $\textbf{P}$ and $\textbf{E}^{AP}$. In general, because matrix $\textbf{C}^{-1}$ is full, we would like to replace it by a block-diagonal matrix $\textbf{A}_{eff}$ that provides the same solution for $\textbf{P}$, given the same initial conditions for $\textbf{E}^{AP}$:
\begin{equation}
    \textbf{P} = \textbf{C}^{-1} \textbf{E}^{AP} \equiv \textbf{A}_{eff}\textbf{E}^{AP},
    \label{Eq:EqM25}
\end{equation}
i.e. matrix $\textbf{A}_{eff}$ contains elements $A_{eff,uv} = \alpha_{eff,u}^{MIMO}\delta_{uv}$, which are the effective (or collective) polarizabilities $\alpha_{eff}^{MIMO}$. These are renormalized polarizabilities that correspond to effectively uncoupled elements that produce the same far-field. \\
\indent While it is always possible to determine $\alpha_{eff,u}^{MIMO}$ numerically using Eq.\@(\ref{Eq:EqM25}), in certain cases it is possible to obtain an analytical form. For example, for weak interactions, by assuming nearest-neighbor coupling with the same coupling constant $\kappa_{uv}\equiv \kappa$, we find (see Appendix~\ref{Sec:Appendix_E}):
\begin{equation}
    \alpha_{eff,u}^{MIMO} \approx \frac{\alpha_u^{MIMO}}{1+2\alpha_u^{MIMO} \kappa \left(1+\cos(k_0l_x\sin{\theta_r}) \right)}.
    \label{Eq:EqM26}    
\end{equation}
\noindent Note that, in the absence of coupling, i.e. if $\kappa=0$, we find $\textbf{C} = \textbf{C}_{self}$ and the solution of Eq.\@(\ref{Eq:EqM25}) reduces to the uncoupled polarizability, i.e. $\alpha_{eff,u}^{MIMO} \equiv \alpha_u^{MIMO}$. \\
\indent To demonstrate the impact of coupling on the received power we will now use Eq.\@(\ref{Eq:EqM16}) for the uncoupled polarizability $\alpha_u^{MIMO}$, which is Eq.\@(\ref{Eq:EqM3}) corrected with $e_0$ given by Eq.\@(\ref{Eq:EqM19}), and we will apply the derived polarizability Eq.\@(\ref{Eq:EqM26}) directly into Eq.\@(\ref{Eq:EqM2}). To examine the validity of our approximation we will use the same $\alpha_u^{MIMO}$ to also solve the full system of Eq.\@(\ref{Eq:EqM24}).
The coupling constant is in general distance-dependent and its functional form depends on the specific element implementation (e.g. cut wires, rectangular patches, split ring resonators, etc). Therefore, to determine the coupling constant, full-wave numerical simulations are necessary. However, for electrically small RIS elements the fields from one element interact with neighboring elements that are almost point-like and, in this case, the coupling constant can be introduced analytically (e.g. see \cite{TretyakovBOOK, Niemi2003}). Here we will use the analytical form that has been used previously within the Discrete Dipole Approximation (DDA) technique \cite{Draine1994} (see Appendix~\ref{Sec:Appendix_E}):
\begin{equation}
    \kappa = C_{\kappa} \kappa_0 \equiv C_{\kappa} \frac{1}{4\pi \epsilon_0} \frac{e^{-j k_0 d}}{d} \left(k_0^2 - \frac{j k_0 d+1}{d^2} \right),
    \label{Eq:EqM27}  
\end{equation}
where $\kappa_0$ is the coupling constant for electric dipoles distributed on the $xz$-plane with periodicity $d\equiv l_x=l_y$, and we have introduced the dimensionless parameter $C_{\kappa}$, which we tune within the range $[0 \ldots 1$] in order to externally control the coupling strength. \\
%
%
%
\begin{figure}[t!]
\centering
		\includegraphics[width=1\linewidth]{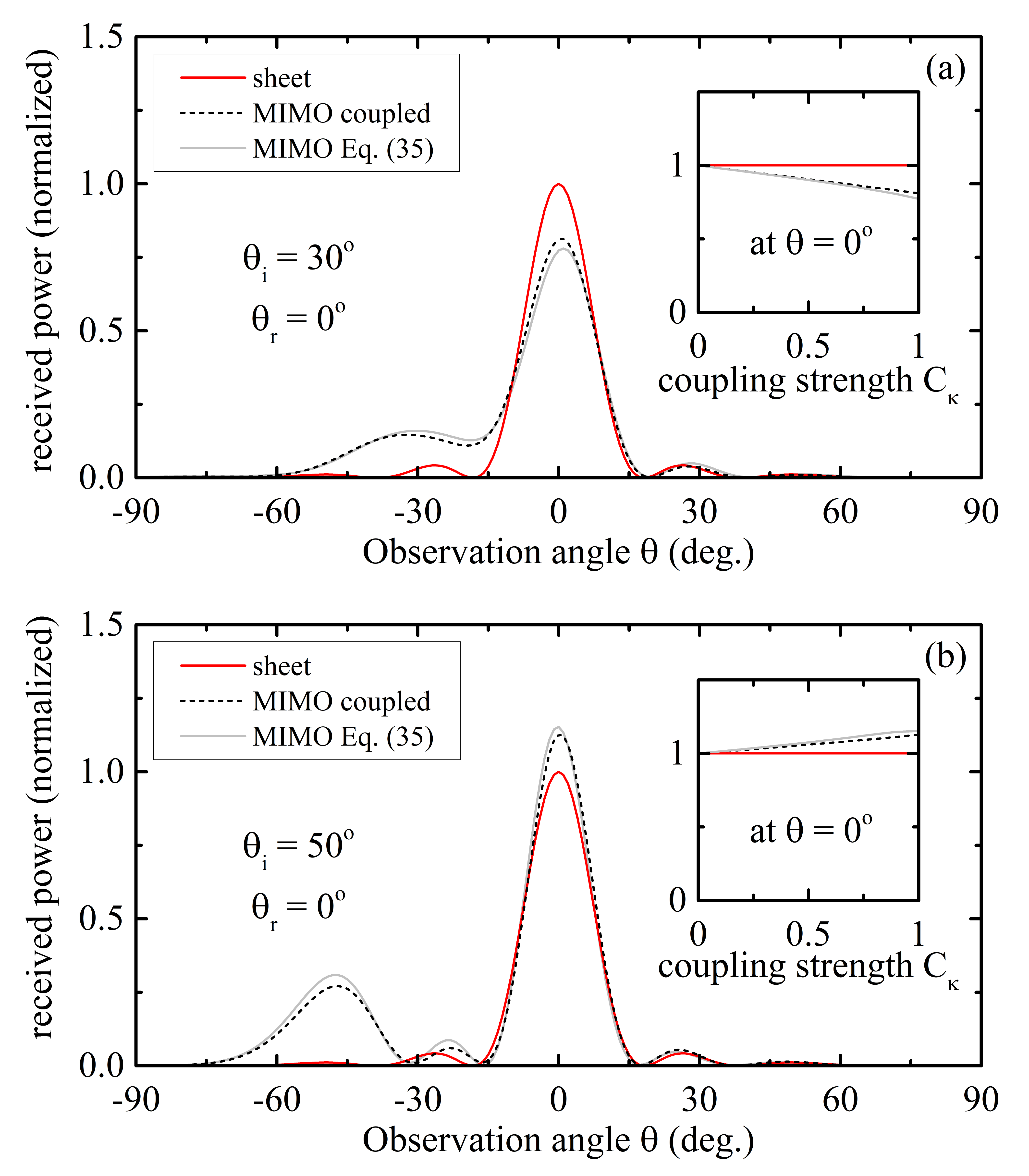}
	\caption{Impact of coupling for RIS elements modeled as antennas with radiation pattern $U_0(\theta,\phi)=\cos^2{\theta}$ and properties consistent with antenna theory. The beam is incident at (a) $\theta_i=30^o$ and (b) $\theta_i=50^o$, in both cases steered towards $\theta_r=0^o$. Depending on the angle of incidence, element coupling may lead to reduced or increased power at the receiver and the excitation of undesired side lobes. The RIS is operating at 20 GHz and consists of $16\times 16$ elements with periodicity $\lambda_0/5$ along both $x$ and $y$ directions, forming a reflecting surface of total size 4.8 cm $\times$ 4.8 cm. The results have been normalized to the maximum received power for the sheet approach or, equivalently, for the uncoupled MIMO model using the correction factor.}
    	\label{fig:fig10}
\end{figure}
%
\indent In Fig.\,\ref{fig:fig10} we study a RIS operating at 20 GHz ($\lambda_0=15$ mm) with $16\times 16$ elements and periodicity $\lambda_0/5$ (size 4.8 cm $\times$ 4.8 cm). The incident beam illuminates the RIS at (a) $\theta_i=30^o$ and (b) $\theta_i=50^o$, and the RIS steers the beam towards $\theta_r=0^o$. The solid red lines show the received power for the sheet model, which is equivalent to the uncoupled MIMO model using the correction factor. When the MIMO elements are coupled (dashed lines) we observe that, depending on the angle of incidence, element coupling may lead to either reduced or increased power at the receiver and the excitation of undesired side lobes. In the insets we monitor the received power at the steering angle as a function of the coupling strength. Our analytical model for the renormalized polarizability under nearest-neighbor coupling (solid gray lines) shows good agreement with the full matrix problem of Eq.\@(\ref{Eq:EqM24}) that we solve self-consistently, and its validity extends to a wide range of steering angles, beyond the examples demonstrated here.\\
\indent The observed deviations in the received power are a consequence of the change of the element polarizability due to coupling. By requiring $\alpha^{MIMO}_{eff,n_x,n_y} = \alpha^{sheet}_{n_x,n_y}$, we find the new correction factor under coupling, $e_0'$, expressed in terms of $e_0$ as (see Appendix~\ref{Sec:Appendix_F} for derivation):
\begin{equation}
    e_0'(n_x,n_y) = e_0 \frac{R_{n_x,n_y}}{R_{n_x,n_y}'},
    \label{Eq:EqM28}    
\end{equation}
where
\begin{equation}
    R_{n_x,n_y}' = \frac{R_{n_x,n_y}}{1+2 R_{n_x,n_y} \kappa \left(1+\cos(k_0l_x\sin{\theta_r}) \right) \frac{2\pi\epsilon_0}{k_0^3} D_{UC} \sqrt{U_{UC}^{AP} U_{UC}^{UE}}}.
    \label{Eq:EqM29}  
\end{equation}
Evidently, under coupling, the correction factor loses its global character and becomes a position-dependent, complex parameter. This local dependence expresses the modification of the element polarizabilities due to coupling, which can be equivalently expressed via the modified reflection coefficient $R_{n_x,n_y}'$. As a result, the ratio of the received power between the MIMO and sheet departs from the value of $1/e_0^2$ found in the uncoupled case, and becomes (see Appendix~\ref{Sec:Appendix_F}):
\begin{align}
      \frac{P_{r}^{MIMO}}{P_{r}^{sheet}} = \frac{1}{e_0^2} \frac{\left|\sum_{n_x}\sum_{n_y} h_{n_x,n_y}^{UE} R_{n_x,n_y}' E_{n_x,n_y}^{AP} \right|^2}{\left|\sum_{n_x}\sum_{n_y} h_{n_x,n_y}^{UE} R_{n_x,n_y} E_{n_x,n_y}^{AP} \right|^2}.
     \label{Eq:EqM30}
\end{align}
Note that, while the correction factor does not depend on the AP-RIS and RIS-UE distances (it is associated with the RIS element properties), the ratio $P_{r}^{MIMO}/P_{r}^{sheet}$ involves the free-space path loss associated with the AP-RIS and RIS-UE distances, $h_{n_x,n_y}^{AP}$ and $h_{n_x,n_y}^{UE}$, respectively ($E_{n_x,n_y}^{AP} = h_{n_x,n_y}^{AP} E^{AP}$, see section II.A). However, in the far-field of the AP and RIS where $d^{AP}_{n_x,n_y}\approx d_{AP}$, and $d^{UE}_{n_x,n_y}\approx d_{UE}$, respectively, the terms $h_{n_x,n_y}^{AP}$ and $h_{n_x,n_y}^{UE}$ are eliminated and the ratio $P_{r}^{MIMO}/P_{r}^{sheet}$ becomes independent of the AP-RIS and RIS-UE distances. In the uncoupled limit where $\kappa \rightarrow 0$ and, consequently, $R_{n_x,n_y}' \rightarrow R_{n_x,n_y}$, the value $1/e_0^2$ is restored.
\section{Discussion and conclusion}
\noindent Designs based on continuous sheet models provide the necessary surface impedance for designing RIS elements to steer the wave to a prescribed direction. However, in view of the various ways of implementing realistic RIS elements, the question that is naturally raised is how a specific design performs for different steering angles and whether certain designs promote the scattered waves towards certain directions. Here, the study of $e_0$ essentially attempts to provide the necessary insight, which is important for assessing the RIS efficiency and utilization in realistic indoor or outdoor scenarios. 
In general, the design and optimization of RIS elements requires full wave simulations that can provide the exact structure of the element radiation patterns and that can take into account all details that cannot be captured by the simplified antenna models. For example, in the analysis presented herein it has been implied that only the phase of the scatterers is externally controlled, i.e. that it is possible to tune the phases of the scatterers without affecting the amplitude of their response. In practice, because the scatterer design usually involves resonant modes, by tuning the phases of the RIS elements it is also possible that their amplitudes are affected as well \cite{Yang2016, Dai2020, Bjornson2021, DeRosny2021}. This dependence may have implications on the RIS performance, which may become further complicated if coupling between the RIS elements is also present. \\
\indent In this work, we analyzed the RIS operation under two frequently utilized approaches, where scattering from the RIS results from either the collective excitation of local (discrete) scatterers or from the global response of a continuous radiating surface. We demonstrated the equivalence between the two approaches, and we discussed models commonly used in recent theoretical works. By using point scatterers with properties consistent with the antenna theory we showed how the shape of the RIS element radiation pattern has implications on the discrepancy in the received power between the two approaches, which was found to depend on the steering angle. Overall, we found that the treatment of the RIS as point scatterers may overestimate the scattered field and, therefore, a correction factor must be taken into account, which we calculated analytically. With our work we aim to bridge RIS approaches that have different origin, i.e. antenna theory vs. effective medium approach, and to provide insight into the possible observed discrepancies between the theoretical models, the understanding of which is crucial for assessing the RIS efficiency and for enabling the application of well-known techniques from MIMO models to RIS-aided links.
\section*{Acknowledgments}
This work has received funding from the European Commission's Horizon 2020 research and innovation programme ARIADNE under grant agreement No. $871464$.
\appendix
\section{RIS as MIMO}
\label{Sec:Appendix_A}
\noindent In this section, we derive the result of Eq.\@(\ref{Eq:EqM2}). Let us consider a transmitter (AP) at distance $d_{n_x,n_y}^{AP}$ from the $(n_x,n_y)$ RIS element. The power density at the ($n_x,n_y$) RIS element is:
\begin{align}
    S_{UC} = \frac{|E_{n_x,n_y}^{AP}|^2}{2Z_0},
     \label{Eq:EqA1}
\end{align}
where $E_{n_x,n_y}^{AP}$ is the field amplitude at the position of the ($n_x,n_y$) element from the transmitter. The power received by the RIS element is $P_{UC}^{inc} = S_{UC} A_{UC} U_{UC}^{AP}$ and the power subsequently re-radiated is $P_{UC}^{rfl} = |R_{n_x,n_y}|^2 P_{UC}^{inc} = |R_{n_x,n_y}|^2 S_{UC} A_{UC} U_{UC}^{AP}$. The power density at the receiver (UE), which is at distance $d_{n_x,n_y}^{UE}$ from the ($n_x,n_y$) RIS element, is:
\begin{multline}
    S_{r} = G_{UC} U_{UC}^{UE} \frac{P_{UC}^{rfl}}{4\pi (d_{n_x,n_y}^{UE})^2} = \\ = G_{UC} U_{UC}^{UE} \frac{|R_{n_x,n_y}|^2 S_{UC} A_{UC} U_{UC}^{AP}}{4\pi (d_{n_x,n_y}^{UE})^2}.
     \label{Eq:EqA2}
\end{multline}
Using Eqs.\@(\ref{Eq:EqA1}),\@(\ref{Eq:EqA2}) to calculate $\sqrt{2Z_0S_r}$, the magnitude of the $E-$field reaching the receiver, and taking into account the accumulated phase, the total field is written as \cite{DiRenzo2021, Alexiou2021}:
\begin{multline}
    E_{n_x,n_y}^{UE} = \sqrt{2Z_0S_r}e^{-j k_0 d_{n_x,n_y}^{UE}} = \\ = R_{n_x,n_y} \sqrt{A_{UC} U_{UC}^{AP} G_{UC} U_{UC}^{UE}} \frac{e^{-j k_0 d_{n_x,n_y}^{UE}}}{\sqrt{4\pi} d_{n_x,n_y}^{UE}}E_{n_x,n_y}^{AP}.
     \label{Eq:EqA3}
\end{multline}
\section{RIS as radiating sheet}
\label{Sec:Appendix_B}
\noindent \noindent In this section, we derive the received power for the radiatng sheet. The directions of incidence and reflection are defined by the wavevectors \textbf{k}$_i$ and \textbf{k}$_r$, respectively, which for plane waves traveling on the $xz$-plane are expressed with respect to the elevation ($\theta$) and azimuth ($\phi$) angles of incidence (subscript $i$) and reflection (subscript $r$) as (see Fig.\,\ref{fig:fig03}):
\begin{subequations}
\begin{gather}
        \textbf{k}_i = k_0 (-\sin{\theta_i}\cos{\phi_i}\hat{\textbf{x}} +\sin{\theta_i}\sin{\phi_i}\hat{\textbf{y}}-\cos{\theta_i}\hat{\textbf{z}}), \\
        \textbf{k}_r = k_0 (\sin{\theta_r}\cos{\phi_r}\hat{\textbf{x}} +\sin{\theta_r}\sin{\phi_r}\hat{\textbf{y}}+\cos{\theta_r}\hat{\textbf{z}}),
\end{gather}
    \label{Eq:EqB1}
\end{subequations}
For simplicity we consider plane waves traveling on the $xz$-plane, i.e. with $\phi_i=\pi$, $\phi_r=0$, and hence the above wavevectors become:
\begin{subequations}
\begin{gather}
        \textbf{k}_i = k_0 (\sin{\theta_i}\hat{\textbf{x}}-\cos{\theta_i}\hat{\textbf{z}}), \\
        \textbf{k}_r = k_0 (\sin{\theta_r}\hat{\textbf{x}}+\cos{\theta_r}\hat{\textbf{z}}).
\end{gather}
    \label{Eq:EqB2}
\end{subequations}
The incident wave can be always expressed as a superposition of TE ($H_x, E_y, H_z$ components) and TM ($E_x, H_y, E_z$ components) polarizations. Here we will examine TE waves and the analysis for TM waves follows similar steps.
\noindent For TE-polarization the incident wave is written as:
\begin{subequations}
\begin{gather}
        \textbf{E}_i(\textbf{r}) = E_i e^{-j \textbf{k}_i \textbf{r}} \hat{\textbf{y}}, \\
        \textbf{H}_i(\textbf{r}) = \frac{E_i}{Z_0} e^{-j \textbf{k}_i \textbf{r}} (\cos{\theta_i} \hat{\textbf{x}}+\sin{\theta_i} \hat{\textbf{z}}),
\end{gather}
    \label{Eq:EqB3}
\end{subequations}
where $\textbf{r} = x \hat{\textbf{x}} + y \hat{\textbf{y}} + z \hat{\textbf{z}}$ is the observation vector and the triplet ($\textbf{E}_i, \textbf{H}_i, \textbf{k}_i$) forms a right-handed orthogonal system, in compliance with Maxwell's equations. Assuming that the sheet preserves the incident polarization, the reflected wave is also TE-polarized and is similarly expressed as:
\begin{subequations}
\begin{gather}
        \textbf{E}_r(\textbf{r}) = E_r e^{-j \textbf{k}_r \textbf{r}} \hat{\textbf{y}} \equiv \Gamma_0 E_i e^{-j \textbf{k}_r \textbf{r}} \hat{\textbf{y}}, \\
        \textbf{H}_r(\textbf{r}) = \frac{\Gamma_0 E_i}{Z_0} e^{-j \textbf{k}_r \textbf{r}} (-\cos{\theta_r} \hat{\textbf{x}}+\sin{\theta_r} \hat{\textbf{z}}),
\end{gather}
    \label{Eq:EqB4}
\end{subequations}
where $\Gamma_0$ is a complex constant associating the incident with the reflected wave amplitude as $E_r=\Gamma_0 E_i$. The solution of the boundary conditions at $z=0$ yields the \textit{surface reflection coefficient}, which is given by \cite{Tretyakov2021}:
\begin{align}
    \Gamma_s(x) \equiv \Gamma_0 e^{j k_0(\sin{\theta_i}-\sin{\theta_r})x},
        \label{Eq:EqB5}
\end{align}
 and guarantees perfect transformation of a single incident plane wave to a single reflected plane wave, i.e. $\textbf{E}_r(\textbf{r})=\Gamma_s \textbf{E}_i(\textbf{r})$ at $z=0$, for an infinitely large surface. \\
\indent For finite-sized sheets, to determine the scattered field of a radiating sheet of size $L_x \times L_y$ as shown in Fig.\,\ref{fig:fig03}, let us consider a vector $\textbf{r}' = x' \hat{\textbf{x}} + y' \hat{\textbf{y}}$ confined on the sheet surface $S$, and also express the observation vector $\textbf{r}$ in spherical coordinates as $\textbf{r} = r\sin{\theta}\cos{\phi}\hat{\textbf{x}} + r\sin{\theta}\sin{\phi}\hat{\textbf{y}} + r\cos{\theta}\hat{\textbf{z}}$ ($\theta$: elevation angle, $\phi$: azimuth angle), where $r=|\textbf{r}|=\sqrt{x^2+y^2+z^2}$ and $\hat{\textbf{r}} = \textbf{r}/r$ is the unit vector. If the surface electric and magnetic currents of the sheet, $\textbf{j}_{se}$ and $\textbf{j}_{sm}$ respectively, are known, then the fields reflected from the sheet at $\textbf{r}$ can be expressed in terms of electric and magnetic vector potentials, $\textbf{A}_{se}$ and $\textbf{A}_{sm}$ respectively, as \cite{OrfanidisBOOK} (Eq. 18.2.8):
\begin{subequations}
\begin{gather}
    \textbf{E}_r(\textbf{r}) = \frac{1}{j\omega \epsilon_0 \mu_0}(\nabla \times \nabla \times \textbf{A}_{se}(\textbf{r})) - \frac{1}{\epsilon_0}\nabla \times \textbf{A}_{sm}(\textbf{r}), \\
    \textbf{H}_r(\textbf{r}) = \frac{1}{j\omega \epsilon_0 \mu_0}(\nabla \times \nabla \times \textbf{A}_{sm}(\textbf{r})) + \frac{1}{\mu_0}\nabla \times \textbf{A}_{se}(\textbf{r}),
\end{gather}
    \label{Eq:EqB6}
\end{subequations}
where 
\begin{subequations}
\begin{gather}
    \textbf{A}_{se}(\textbf{r}) = \iint_{S} \mu_0 \textbf{j}_{se}(\textbf{r}') G(\textbf{r}-\textbf{r}')d\textbf{r}', \\
    \textbf{A}_{sm}(\textbf{r}) = \iint_{S} \epsilon_0 \textbf{j}_{sm}(\textbf{r}') G(\textbf{r}-\textbf{r}')d\textbf{r}',
\end{gather}
    \label{Eq:EqB7}
\end{subequations}
and
\begin{align}
        G(\textbf{r}-\textbf{r}') = \frac{e^{-j k_0 |\textbf{r}-\textbf{r}'|}}{4\pi |\textbf{r}-\textbf{r}'|}
        \label{Eq:EqB8}
\end{align}
is the Green’s function for the Helmholtz equation, $\omega$ is the angular frequency, and $\epsilon_0, \mu_0$ are the vacuum permittivity and permeability, respectively. For surfaces that are significantly larger than $\lambda_0$, so that the edge effects can be neglected, as is usually the case in practical situations, we may apply the boundary conditions at $z=0$ and use the incident and reflected fields [Eqs.\@(\ref{Eq:EqB3}),\@(\ref{Eq:EqB4})] to replace the surface currents with \cite{SurfaceEMBOOK}:
\begin{subequations}
    \begin{gather}
        \textbf{j}_{se} = +\hat{\textbf{z}}\times (\textbf{H}_i(\textbf{r}')+\textbf{H}_r(\textbf{r}')), \\
        \textbf{j}_{sm} = -\hat{\textbf{z}}\times (\textbf{E}_i(\textbf{r}')+\textbf{E}_r(\textbf{r}')).
    \end{gather}
    \label{Eq:EqB9}
\end{subequations}
\noindent For simplicity, we will treat here the sheet as a radiating aperture, which corresponds to omitting the incident field from the boundary conditions \@(\ref{Eq:EqB9}). In this case, using only Eq.\@(\ref{Eq:EqB4}), we find:
\begin{subequations}
\begin{gather}
    \textbf{j}_{se} = \frac{\cos{\theta_r}}{Z_0}\Gamma_0 E_i \hat{\textbf{y}}, \\
    \textbf{j}_{sm} = -\Gamma_0 E_i \hat{\textbf{x}}.
\end{gather}
    \label{Eq:EqB10}
\end{subequations}
\noindent For distances $|\textbf{r}-\textbf{r}'| \gg \lambda_0$, we may approximate:
\begin{subequations}
\begin{multline}
    \nabla \times \left(\textbf{j}_{\{se,sm\}}(\textbf{r}') G(\textbf{r}_s)\right) \approx \\ -j k_0 G(\textbf{r}_s) \left(\hat{\textbf{r}}_s \times \textbf{j}_{\{se,sm\}}(\textbf{r}')\right),
\end{multline}
\begin{multline}
    \nabla \times \nabla \times \left(\textbf{j}_{\{se,sm\}}(\textbf{r}') G(\textbf{r}_s)\right) \approx \\ -k_0^2 G(\textbf{r}_s) \left(\hat{\textbf{r}}_s \times \hat{\textbf{r}}_s \times \textbf{j}_{\{se,sm\}}(\textbf{r}')\right),
\end{multline}
    \label{Eq:EqB11}
\end{subequations}
where $\textbf{r}_s \equiv \textbf{r}-\textbf{r}'$. Inserting Eq.\@(\ref{Eq:EqB10}) into Eqs.\@(\ref{Eq:EqB6}), (\ref{Eq:EqB7}) and using Eq.\@(\ref{Eq:EqB11}), we find:
\begin{subequations}
    \begin{gather}
        \textbf{E}_r(\textbf{r}) = j k_0 \iint_{S} E_r(\textbf{r}') G(\textbf{r}-\textbf{r}') \textbf{V}_e(\textbf{r}-\textbf{r}') d\textbf{r}', \\
        \textbf{H}_r(\textbf{r}) = \frac{j k_0}{Z_0} \iint_{S} E_r(\textbf{r}') G(\textbf{r}-\textbf{r}') \textbf{V}_m(\textbf{r}-\textbf{r}')  d\textbf{r}',
    \end{gather}
    \label{Eq:EqB12}
\end{subequations}
where $E_r(\textbf{r}') =  \Gamma_s(\textbf{r}') E_i(\textbf{r}')$ and the vectors $\textbf{V}_e$ and $\textbf{V}_m$ are explicitly written as:
\begin{subequations}
\begin{gather}
     \textbf{V}_e = \cos{\theta_r}\left(\hat{\textbf{r}}_s \times \hat{\textbf{r}}_s \times \textbf{y}\right) - \left(\hat{\textbf{r}}_s \times \textbf{x}\right) \\
     \textbf{V}_m = -\left(\hat{\textbf{r}}_s \times \hat{\textbf{r}}_s \times \textbf{x}\right) - \cos{\theta_r}\left(\hat{\textbf{r}}_s \times \textbf{y}\right),
\end{gather}
    \label{Eq:EqB13}
\end{subequations}
The power density at $\textbf{r}$ is given by the Poynting vector, i.e. $\textbf{S}_r(\textbf{r})=\frac{1}{2}$ Re$(\textbf{E}_r \times \textbf{H}^*_r)$, and involves the calculation of the integrals in Eq.\@(\ref{Eq:EqB12}). In the far-field, i.e. approximately after distance $2\times$\,max$(L_x^2,L_y^2)/\lambda_0$, where $\textbf{V}_e(\textbf{r}-\textbf{r}') \approx \textbf{V}_e(\textbf{r})$, $\textbf{V}_m(\textbf{r}-\textbf{r}') \approx \textbf{V}_m(\textbf{r})$, we find:
\begin{align}
    \textbf{S}_r(\textbf{r}) = \frac{k_0^2}{2Z_0} \Theta_r(\textbf{r})\left|I_r\right|^2 \hat{\textbf{r}},
    \label{Eq:EqB14}
\end{align}
where 
\begin{multline}
    \Theta_r(\textbf{r}) \equiv |\textbf{V}_e \times \textbf{V}_m|= \\ = \frac{y^2}{r^2} + \frac{z^2}{r^2} + \frac{2z}{r} \cos{\theta_r} + \left(\frac{x^2}{r^2} + \frac{z^2}{r^2}\right) \cos^2{\theta_r},
    \label{Eq:EqB15}
\end{multline}
and
\begin{align}
    I_r = \iint_{S} E_r(\textbf{r}') G(\textbf{r}-\textbf{r}')d\textbf{r}'.
    \label{Eq:EqB16}  
\end{align}
In spherical coordinates $\Theta_r$ takes the form of Eq.\@(\ref{Eq:EqM6}) and the received power, which is calculated as $P_r = A_r \textbf{S}_r(\textbf{r}) \cdot \hat{\textbf{r}}$, corresponds to the first term in Eq.\@(\ref{Eq:EqM5}), \\
\indent Using the full form of \@(\ref{Eq:EqB9}) leads to the full form of \@(\ref{Eq:EqM5}), the derivation of which follows similar steps and involves the integral $I_i$, which is defined as: 
\begin{align}
    I_i = \iint_{S} E_i(\textbf{r}') G(\textbf{r}-\textbf{r}')d\textbf{r}'.
    \label{Eq:EqB17}  
\end{align}
Note that, the parameter $\Theta_r$, which was found to play a crucial role in the equivalence between the MIMO and sheet models, is identical for TM waves and the observations are therefore general and applicable to any linear polarization.
\section{RIS as radiating aperture}
\label{Sec:Appendix_C}
\noindent To calculate the deviation in the received power when treating the RIS as a radiating aperture, i.e. considering only the $\Theta_r$ term in the power calculations, we start with writing the full expression for the received power of the radiating sheet [Eq.\@(\ref{Eq:EqM5})] as $P_r = P_r^A+P_r^B+P_r^{AB}$, where:
\begin{subequations}
\begin{gather}
    P_r^A(\theta, \phi) = A_r \frac{k_0^2}{2Z_0} \Theta_r \left|I_r\right|^2, \\
    P_r^B(\theta, \phi) = A_r \frac{k_0^2}{2Z_0} \Theta_i \left|I_i\right|^2, \\
    P_r^{AB}(\theta, \phi) = A_r \frac{k_0^2}{2Z_0} \Theta_{ir} 2 Re(I_i I_r^*).
\end{gather}
    \label{Eq:EqC1}  
\end{subequations}
For plane wave illumination the integrals $I_i,I_r$ are calculated analytically as:
\begin{multline}
    I_r = \Gamma_0 E_{i} L_xL_y \frac{e^{-jk_0r}}{4\pi r} \times \\ \frac{\sin{\left(\frac{1}{2}k_0L_x\left(\sin{\theta}\cos{\phi}-\sin{\theta_r}\right)\right)}}{\frac{1}{2}k_0L_x\left(\sin{\theta}\cos{\phi}-\sin{\theta_r}\right)} \times\frac{\sin{\left(\frac{1}{2}k_0L_y\sin{\theta}\sin{\phi}\right)}}{\frac{1}{2}k_0L_y\sin{\theta}\sin{\phi}}. 
    \label{Eq:EqC2}      
\end{multline}
\begin{multline}
    I_i = E_{i} L_xL_y \frac{e^{-jk_0r}}{4\pi r} \times \\ \frac{\sin{\left(\frac{1}{2}k_0L_x\left(\sin{\theta}\cos{\phi}-\sin{\theta_i}\right)\right)}}{\frac{1}{2}k_0L_x\left(\sin{\theta}\cos{\phi}-\sin{\theta_i}\right)} \times\frac{\sin{\left(\frac{1}{2}k_0L_y\sin{\theta}\sin{\phi}\right)}}{\frac{1}{2}k_0L_y\sin{\theta}\sin{\phi}}. 
    \label{Eq:EqC3}      
\end{multline}
At the position of the receiver, where $\theta=\theta_r, \phi=0$, we find that:
\begin{align}
    \frac{P_r^B(\theta_r,0)+P_r^{AB}(\theta_r,0)}{P_r^A(\theta_r,0)}=\left(\frac{X}{|\Gamma_0|}\right)^2+2\frac{X}{|\Gamma_0|},
    \label{Eq:EqC4}      
\end{align}
where $|\Gamma_0|$ is the magnitude of the reflection coefficient [see Eq.\@(\ref{Eq:EqB5})] and:
\begin{align}
    X = \frac{1}{2}\left(1-\frac{\cos\theta_i}{\cos\theta_r}\right)\frac{\sin{\left(\frac{k_0L_x}{2}\left(\sin{\theta_i}-\sin{\theta_r}\right)\right)}}{\frac{k_0L_x}{2}\left(\sin{\theta_i}-\sin{\theta_r}\right)}.
    \label{Eq:EqC5}      
\end{align}
For the RIS studied in this work, i.e. consisting of $250\times 250$ elements with size $l_x=l_y=\lambda_0/5$, we find that $X<1\%$, except for extreme incidence and reflection close to $90^o$. Note that as the RIS size increases, i.e. as more elements of the same size are added, this fraction is further suppressed. Exacty at specular reflection, it is entirely eliminated for any RIS size.
\section{Discretizing the continuous sheet}
\label{Sec:Appendix_D}
\noindent To calculate the effective dipole polarizability $\alpha_{n_x,n_y}^{sheet}$ of the discretized radiating sheet we first split the integrals in Eq.\@(\ref{Eq:EqB7}) into sums of integrals within each small rectangle of size $l_x\times l_y$, which we then substitute with their averaged values:
\begin{subequations}
\begin{multline}
    \iint_{S} \mu_0 \textbf{j}_{se} G(\textbf{r}-\textbf{r}')d\textbf{r}' = \sum_{n_x,n_y} \iint_{S_{n_x,n_y}} \mu_0 \textbf{j}_{se} G(\textbf{r}-\textbf{r}')d\textbf{r}' \\ \approx  l_xl_y\sum_{n_x,n_y} \mu_0 \textbf{j}_{se}^{n_x,n_y} G_{n_x,n_y}, 
\end{multline}
\begin{multline}
    \iint_{S} \epsilon_0 \textbf{j}_{sm} G(\textbf{r}-\textbf{r}')d\textbf{r}' = \sum_{n_x,n_y} \iint_{S_{n_x,n_y}} \epsilon_0 \textbf{j}_{sm} G(\textbf{r}-\textbf{r}')d\textbf{r}' \\ \approx  l_xl_y\sum_{n_x,n_y} \epsilon_0 \textbf{j}_{sm}^{n_x,n_y} G_{n_x,n_y},
\end{multline}
    \label{Eq:EqD1}
\end{subequations}
Here $\textbf{j}_{se}^{n_x,n_y}, \textbf{j}_{sm}^{n_x,n_y}$ are the surface currents and $G_{n_x,n_y}$ is the Green's function, all calculated at the position of the ($n_x,n_y$) dipole; $S_{n_x,n_y}$ denotes the area of the ($n_x,n_y$) rectangle. To relate the discretized surface currents with the dipole polarizabilities, we can now write the former as $\textbf{j}_{se}^{n_x,n_y} = (j\omega/l_xl_y)\textbf{p}_{n_x,n_y}$, $\textbf{j}_{sm}^{n_x,n_y} = (j\omega/l_xl_y) \textbf{m}_{n_x,n_y}$ \cite{TretyakovBOOK, SurfaceEMBOOK}, where $\textbf{p}_{n_x,n_y}, \textbf{m}_{n_x,n_y}$ are the electric and magnetic dipole moments of the ($n_x,n_y$) element, respectively. We obtain:
\begin{subequations}
\begin{gather}
    \textbf{A}_{se} \approx j\omega \mu_0 \sum_{n_x,n_y} \textbf{p}_{n_x,n_y} G_{n_x,n_y}, \\
    \textbf{A}_{sm} \approx j\omega \epsilon_0 \sum_{n_x,n_y} \textbf{m}_{n_x,n_y} G_{n_x,n_y}.
\end{gather}
    \label{Eq:EqD2}
\end{subequations}
Inserting the above result in Eq.\@(\ref{Eq:EqB6}a), and with the aid of Eq.\@(\ref{Eq:EqB11}), we reach the result of Eq.\@(\ref{Eq:EqM10}) for the electric field of the ($n_x,n_y$) dipole. \\
\indent The discretization error in Eq.\@(\ref{Eq:EqD1}) can be lifted if we perform the integrations within each small rectangle, instead of replacing the discrete integrals with their averaged values. In this case the discretized sheet polarizability reads:
\begin{align}
    \alpha_{n_x,n_y}^{sheet} = l_x l_y\frac{j \epsilon_0}{k_0} \Gamma_{n_x,n_y} C\sqrt{\Theta_r(\theta, \phi)}
    \label{Eq:EqD3}
\end{align}
where
\begin{equation}
    C = \frac{\sin{\frac{k_0 l_x}{2}(\sin{\theta}\cos{\phi}-\sin{\theta_r})}}{\frac{k_0 l_x}{2}(\sin{\theta}\cos{\phi}-\sin{\theta_r})} \times\frac{\sin{\frac{k_0 l_y}{2}\sin{\theta}\sin{\phi}}}{\frac{k_0 l_y}{2}\sin{\theta}\sin{\phi}}.
    \label{Eq:EqD4}    
\end{equation}
The correction factor now reads:
\begin{equation}
    e_0 = \frac{4\pi}{D_{UC}}\frac{l_xl_y}{\lambda_0^2} C\sqrt{\frac{\Theta_r(\theta,\phi)}{4U_{UC}^{AP} U_{UC}^{UE}}}.
    \label{Eq:EqD5}
\end{equation}
For plane wave illumination, Eq.\@(\ref{Eq:EqM4}) with the correction factor of Eq.\@(\ref{Eq:EqD5}) reproduces exactly the power predicted by Eq.\@(\ref{Eq:EqM9}) for point scatterers with $|\Gamma_{n_x,n_y}| = \Gamma_0$, regardless of the discretization density. This is guaranteed by the analytically calculated factor $C$ given by Eq.\@(\ref{Eq:EqD4}), which results from the condition that $E_i$ has constant magnitude across each rectangle. Note that $C\rightarrow1$ as the element size is reduced, leading to Eq.\@(\ref{Eq:EqD3})$\rightarrow$Eq.\@(\ref{Eq:EqM16}) and  Eq.\@(\ref{Eq:EqD5})$\rightarrow$Eq.\@(\ref{Eq:EqM19}).
\section{Coupling between RIS elements}
\label{Sec:Appendix_E}
\subsection{Renormalized polarizabilities for nearest-neighbor coupling}
Let us assume that the ($n_x,n_y$) element couples only to the elements with indices ($n_x-1,n_y$), ($n_x+1,n_y$), ($n_x,n_y-1$) and ($n_x,n_y+1$). Using Eq.\@(\ref{Eq:EqM24}) we may write $\textbf{p}_{n_x,n_y} = \alpha_{eff,n_x,n_y}  \textbf{E}^{AP}_{n_x,n_y}$, where:
\begin{equation}
    \alpha_{eff,n_x,n_y} = \frac{\alpha_{n_x,n_y}}{1+\alpha_{n_x,n_y} \kappa\frac{\textbf{p}_{n_x-1,n_y}+\textbf{p}_{n_x+1,n_y}+\textbf{p}_{n_x,n_y+1}+\textbf{p}_{n_x,n_y+1}}{\textbf{p}_{n_x,n_y}}}.
    \label{Eq:EqE1}        
\end{equation}
This result is exact for nearest-neighbor coupling, however requires the knowledge of the neighbor dipole moments, which depend on the local fields that contain the interactions between elements. To derive a closed form we may assume that, for weak coupling, the polarization averaging in the denominator of Eq.\@(\ref{Eq:EqE1}) is not severely modified if we neglect the interactions and use the incident field only. In this case, and taking into account that the element polarizabilities contain the reflection coefficient $\Gamma_{n_x,n_y}$, we find:
\begin{equation}
    \alpha_{eff,n_x,n_y} \approx \frac{\alpha_{n_x,n_y}}{1+\alpha_{n_x,n_y} \kappa\frac{E^{n_x-1,n_y}_r+E^{n_x+1,n_y}_r+E^{n_x,n_y-1}_r+E^{n_x,n_y+1}_r}{E^{n_x,n_y}_r}},
    \label{Eq:EqE2}       
\end{equation}
where $E_r$ is the reflected field at each element. Using Eq.\@(\ref{Eq:EqB4}a) for $E_r$, we reach the final result:
\begin{equation}
    \alpha_{eff,n_x,n_y} \approx \frac{\alpha_{n_x,n_y}}{1+2\alpha_{n_x,n_y} \kappa \left(1+\cos(k_0l_x\sin{\theta_r}) \right)}.
    \label{Eq:EqE3}    
\end{equation}
\subsection{Coupling constant for electric dipoles}
\noindent In Eq.\@(\ref{Eq:EqM24}) we expressed the electric field at $\textbf{r}_u$ that is due to dipole $\textbf{p}_v$ at location $\textbf{r}_v$ as $-\textbf{K}_{uv} \textbf{p}_v$. Each element $\textbf{K}_{uv}$ is a $3\times3$ matrix and, for electric dipoles, is given analytically by \cite{Draine1994}:
\begin{multline}
    \textbf{K}_{uv} = \frac{1}{4\pi \epsilon_0} \frac{e^{-j k_0 r_{uv}}}{r_{uv}}\times \\ \left[k_0^2(\hat{r}_{uv}\hat{r}_{uv}-\textbf{I}) + \frac{-j k_0 r_{uv}-1}{r_{uv}^2}(3\hat{r}_{uv}\hat{r}_{uv}-\textbf{I}) \right]
    \label{Eq:EqE4}      
\end{multline}
where $r_{uv}\equiv|\textbf{r}_u-\textbf{r}_v|$, $\hat{r}_{uv}\equiv(\textbf{r}_u-\textbf{r}_v)/r_{uv}$, and $\textbf{I}$ is the $3\times3$ identity matrix.
For dipoles oriented along the $y$-axis and periodically arranged on the $xy$-plane with periodicity $l_x=l_y\equiv d$, the electric field $\textbf{K}_{uv}\textbf{p}_v$ has only one nonzero component for interactions along the $x$-axis:
\begin{equation}
    \kappa_{uv}|\textbf{p}_v| \equiv \kappa_0|\textbf{p}_v| = \frac{1}{4\pi \epsilon_0} \frac{e^{-j k_0 d}}{d} \left(k_0^2 - \frac{j k_0 d+1}{d^2} \right)|\textbf{p}_v|,
    \label{Eq:EqE5}      
\end{equation}
In our examples we have used $\kappa_0$ in the above equation as an estimate for the coupling constant. 
\section{Received power ratio between MIMO and sheet approach}
\label{Sec:Appendix_F}
\noindent Using Eq.\@(\ref{Eq:EqM3}) and Eq.\@(\ref{Eq:EqM16}) we have found that the uncoupled MIMO polarizability is expressed in terms of the correction factor as:
\begin{equation}
    \alpha_{n_x,n_y}^{MIMO} = e_0\frac{2 \pi \epsilon_0}{k_0^3} R_{n_x,n_y} D_{UC} \sqrt{U_{UC}^{AP} U_{UC}^{UE}}.
    \label{Eq:EqF1}
\end{equation}
Similarly, under coupling, we may write:
\begin{equation}
    \alpha_{eff,n_x,n_y}^{MIMO} = e_0'\frac{2 \pi \epsilon_0}{k_0^3} R_{n_x,n_y}' D_{UC} \sqrt{U_{UC}^{AP} U_{UC}^{UE}},
    \label{Eq:EqF2}
\end{equation}
where $\alpha_{eff,n_x,n_y}^{MIMO}$ is the new polarizability, accounting for the modified reflection coefficient $R_{n_x,n_y}'$, and $e_0'$ is the new correction factor. By requiring $\alpha_{n_x,n_y}^{MIMO} =\alpha_{eff,n_x,n_y}^{MIMO} \equiv \alpha_{n_x,n_y}^{sheet}$, we reach the result of Eq.\@(\ref{Eq:EqM28}), $e_0' = e_0 R_{n_x,n_y}/R_{n_x,n_y}'$. To analytically express $R_{n_x,n_y}'$ in terms of $R_{n_x,n_y}$, we use Eq.\@(\ref{Eq:EqM26}), in which we insert Eq.\@(\ref{Eq:EqF1}) and Eq.\@(\ref{Eq:EqF2}) uncorrected  (i.e. with $e_0=e_0'=1$), which leads to the result of Eq.\@(\ref{Eq:EqM29}). \\
\indent The receiver power is expressed via the sum of the field contributions from all elements as $P_{r} = (A_r/2Z_0)|\sum_{n_x}\sum_{n_y} E^{UE}_{n_x,n_y}|^2$, where $E_{n_x,n_y}^{UE} = h_{n_x,n_y}^{UE} \alpha_{n_x,n_y} E_{n_x,n_y}^{AP}$ is the electric field at the receiver antenna from the $(n_x,n_y)$ RIS element, which is characterized by the polarizability $\alpha_{n_x,n_y}$. Using Eq.\@(\ref{Eq:EqF2}) uncorrected (i.e. with $e_0'=1$) for $\alpha_{n_x,n_y}^{MIMO}$ and Eq.\@(\ref{Eq:EqM16}) with Eq.\@(\ref{Eq:EqM19}) for $\alpha_{n_x,n_y}^{sheet}$, we find that the ratio of the received power between the MIMO and sheet approach is:
\begin{multline}
      \frac{P_{r}^{MIMO}}{P_{r}^{sheet}} = \frac{\left|\sum_{n_x}\sum_{n_y} h_{n_x,n_y}^{UE} \alpha_{eff,n_x,n_y}^{MIMO} E_{n_x,n_y}^{AP} \right|^2}{\left|\sum_{n_x}\sum_{n_y} h_{n_x,n_y}^{UE} \alpha_{n_x,n_y}^{sheet} E_{n_x,n_y}^{AP} \right|^2} = \\ \frac{1}{e_0^2} \frac{\left|\sum_{n_x}\sum_{n_y} h_{n_x,n_y}^{UE} R_{n_x,n_y}' E_{n_x,n_y}^{AP} \right|^2}{\left|\sum_{n_x}\sum_{n_y} h_{n_x,n_y}^{UE} R_{n_x,n_y} E_{n_x,n_y}^{AP} \right|^2}.
    \label{Eq:EqF3}      
\end{multline}
\bibliography{main}
\end{document}